\documentclass[aps,showpacs,groupedaddress,superscriptaddress,twocolumn,longbibliography]{revtex4-1}

\usepackage{graphicx}
\usepackage{dcolumn}
\usepackage{bm}
\usepackage{physics}
\usepackage{amsmath,amssymb}
\usepackage{xcolor}
\usepackage{hyperref}
\hypersetup{
  breaklinks=true,
  colorlinks=true,
  allcolors=blue,
}

\usepackage{mathrsfs}  

\usepackage{wrapfig}

\newcommand{\ox}[1]{{\color{black}{#1}\normalcolor}}

\begin{document}

\preprint{APS/123-QED}
\title{Dynamics of spin-momentum entanglement from superradiant phase transitions}

\author{Oksana Chelpanova\href{https://orcid.org/0000-0002-1679-1359}{\includegraphics[height=1.7ex]{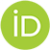}}}
\affiliation{                    
Institut für Physik, Johannes Gutenberg-Universität Mainz, D-55099 Mainz, Deutschland
}
\author{Kushal Seetharam\href{https://orcid.org/0000-0003-3159-8618}{\includegraphics[height=1.7ex]{orcid-logo.png}}}
\affiliation{Department of Electrical Engineering, Massachusetts Institute of Technologies, Cambridge, Massachusetts 02139, USA}
\affiliation{Department of Physics, Harvard University, Cambridge MA, 02138, USA}
\author{Rodrigo Rosa-Medina\href{https://orcid.org/0000-0001-7321-7743}{\includegraphics[height=1.7ex]{orcid-logo.png}}}
\affiliation{Institute for Quantum Electronics, ETH Zürich, 8093 Zürich, Switzerland}
\author{Nicola Reiter\href{https://orcid.org/0009-0004-1830-3779}{\includegraphics[height=1.7ex]{orcid-logo.png}}}
\affiliation{Institute for Quantum Electronics, ETH Zürich, 8093 Zürich, Switzerland}
\author{Fabian Finger\href{https://orcid.org/0000-0002-6996-5957}{\includegraphics[height=1.7ex]{orcid-logo.png}}}
\affiliation{Institute for Quantum Electronics, ETH Zürich, 8093 Zürich, Switzerland}
\author{Tobias Donner\href{https://orcid.org/0000-0001-7016-587X}{\includegraphics[height=1.7ex]{orcid-logo.png}}}
\affiliation{Institute for Quantum Electronics, ETH Zürich, 8093 Zürich, Switzerland}
 \author{Jamir Marino\href{https://orcid.org/0000-0003-2585-2886}{\includegraphics[height=1.7ex]{orcid-logo.png}}}
\affiliation{                    
Institut für Physik, Johannes Gutenberg-Universität Mainz, D-55099 Mainz, Deutschland
}

\date{\today}

\begin{abstract}
Exploring operational regimes of many-body cavity QED with multi-level atoms remains an exciting research frontier for their enhanced storage capabilities of intra-level quantum correlations. 
In this work, we consider an experimentally  feasible many-body cavity QED model describing a four-level system, where each of those levels is formed from a combination of different spin and momentum states of ultra-cold atoms in a cavity.
%
The resulting model comprises a pair of Dicke Hamiltonians constructed from pseudo-spin operators, effectively capturing two intertwined superradiant phase transitions. 
The  phase diagram reveals 
regions featuring weak and strong entangled states of spin and momentum atomic degrees of freedom. These states exhibit different dynamical responses, ranging from slow to fast relaxation, with the added option of persistent entanglement temporal oscillations.
We discuss the role of cavity losses in steering the system's dynamics into such entangled states and propose a readout scheme that leverages different light polarizations within the cavity. 
Our work paves the way to connect the rich variety of non-equilibrium phase transitions that occur in many-body cavity QED to the buildup of quantum correlations in systems with multi-level atom descriptions. 
\end{abstract}

\maketitle

\section{\label{sec:intro}Introduction}

The coupling between spin and motional degrees of freedom lies at the root of several rich phenomena in quantum physics, including 
 fine structure splitting of atoms~\cite{sakurai1995modern}
 and the spin Hall effect~\cite{SpinHall}, which, in turn, allow the realization of topological phases of matter~\cite{TopoIndulators,sato2017topological} and open the possibility of topological quantum computing~\cite{TopQComputation,qi2023spin}.
Spin-momentum entanglement resulting from such coupling has increasingly become relevant in a variety of research areas, ranging from materials science to photonics and atomic systems~\cite{stav2018quantum,kale2020spin,DChang2017designing}.  
In this work, we propose a protocol for engineering entanglement between the spin and momentum degrees of freedom of 
ultracold atoms coupled to an optical cavity. Our approach exploits a non-equilibrium superradiant phase transition in the system realized by  coupling four atomic modes, which comprise two internal (spin) and external (momentum) states of the atom.

\begin{figure}
    \centering
 \includegraphics[width=\linewidth]{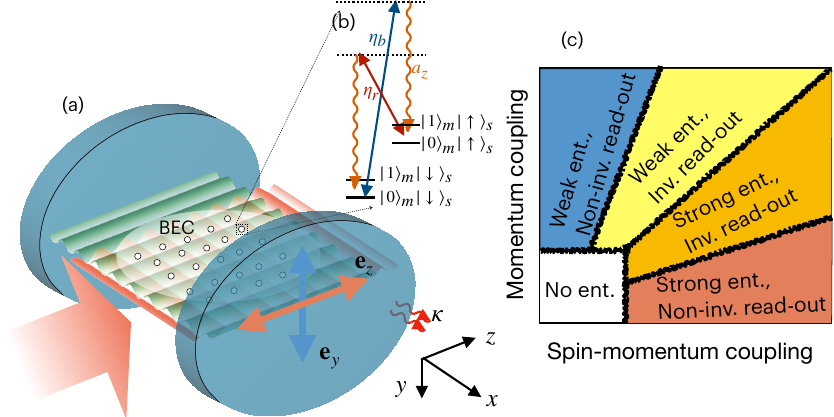}
    \caption{(a) Sketch of the cavity QED setup and (b) of the atomic level scheme. (c) Cartoon of the {dynamical} phase diagram as a function of the  couplings between spin and momentum degrees of freedom. 
    }
    \label{fig:entanglement0}
\end{figure}

Many-body cavity QED experiments with ultracold atoms are among the most versatile quantum simulators of driven-dissipative phases of matter~\cite{mivehvar2021cavity}. 
The combination of tunable photon-mediated long-range interatomic interactions, along with strong cooperative effects and control on cavity losses, offers a wide range of possibilities, encompassing non-equilibrium transitions~\cite{dogra2019dissipation,baumann2010dicke,kroeze2019dynamical,marino2022dynamical,Kessler_open_Dicke_nodel,Kessler1,Kessler2,Phatthamon_continuousTC,AM_BCS,PhysRevLett.126.230601},
dynamical control of  correlations~\cite{Kushal1,Kushal2,Jamir_universality,periwal2021programmable,finger2023spin,RosaMedinaMomentumLattice},
realization of dark states~\cite{Lin2022,DarkStateCond,Orioli_Dark}, and the exploration of collective phenomena purely driven by engineered dissipation  ~\cite{soriente2018dissipation,landini2018formation,Dreon_2022}.
Oftentimes, the effective atomic degrees of freedom in state-of-art experiments are a pair of momentum states or internal levels, optically addressed by external laser drives inducing cavity-assisted two-photon transitions~\cite{NagyDickeModelPhaseTransition,baumann2010dicke,zhiqiang2017nonequilibrium,PhysRevLett.119.213601}.
Recently, a few cavity QED experiments and theory works have shown how to couple  the momentum and internal spin degrees of freedom of ultracold atoms using intra-cavity light, demonstrating novel self-organized    phases ~\cite{Ferri,LevSpinorSelfOrdering, dogra2019dissipation,PhysRevLett.119.063602,Jager_squeezing}. 

In this paper, we generalize such protocols   to show that spin-momentum entanglement can be synthesized, controlled, and steered in
experiments by coupling motional and internal degrees of freedom. 
We consider a cavity QED platform, cf.   Fig.~\ref{fig:entanglement0}(a), which is described by a minimal model with two different spin states and two different momentum states, meaning that each atom can occupy one of these four hybrid spin-momentum states  shown in Fig.~\ref{fig:entanglement0}(b).
We demonstrate that this system exhibits superradiant phase transitions related to the self-organization of the atoms in the cavity, concomitantly with the dynamical buildup of spin-momentum entanglement. 
By varying the spin-momentum coupling, one can robustly tune the entanglement up to its maximum possible value.

The superradiant phase transitions within this model exhibit notable distinctions from the conventional phenomenology of self-organization in cavity  QED
~\cite{dogra2019dissipation,baumann2010dicke,NagyDickeModelPhaseTransition,Nagy_OpenDickeModel,Kevin,Lerose,defenu2023outofequilibrium,kirton2019introduction,landini2018formation,Dreon_2022}. While the hybrid spin-momentum order parameter in our system and the photon number  approach stationary values, the spin and momentum separately can be non-stationary. This results in a time-dependent profile of the condensate density and the effective spin magnetization, which is   unconventional   for state-of-art cavity QED experiments~\cite{Ferri,LevSpinorSelfOrdering}. 
Such oscillations  persist beyond the  operational time scales of these platforms, resulting in long-lived non-stationary dynamical responses. 
We show that such features can be continuously probed using an auxiliary cavity field, which is coupled to the momentum degree of freedom of atoms. The overall dynamics in such a model are conditioned by the intertwining of two cavity-mediated processes, controlled by two couplings between momentum states or hybrid momentum-spin states. The different dynamical responses of the system, summarized in Fig.~\ref{fig:entanglement0}(c), are characterized by weak or strong entanglement. In particular, despite the back-action and intrinsic decoherence of the read-out process,    {proxy of} spin-momentum entanglement 
dynamics
can be non-invasively accessed in an extended parameter regime [red region in Fig.~\ref{fig:entanglement0}(c)], making the system a possible  candidate for quantum information applications~\cite{nielsen_chuang_2010}.

Crucially, in our scheme, 
cavity losses have the beneficial role of steering dynamics towards target entangled states, thereby endowing robustness to the initial condition of the system. This feature is absent in protocols engineering spin-momentum entanglement in BECs using solely classical drive fields~\cite{kale2020spin}. 
In such proposals, the degree of achievable spin-momentum entanglement is  highly sensitive with respect to technical fluctuations of different experimental parameters (e.g., drive powers and frequencies). 
In contrast, protocols relying on cavity losses induce contractive dynamics which are   insensitive to such issues  and initial state preparation,
therefore offering a more robust and reliable route for spin-momentum entanglement generation.

\subsection{Outline of the article}

The   paper is organized as follows.  
In Sec.~\ref{sec:Model},  we present an experimentally motivated effective model that governs the dynamics of the cavity QED setup in Fig.~\ref{fig:entanglement0}, where the spin and momentum of atoms are coupled to the cavity field.
Sec.~\ref{sec:ent_ph_tr_1} is devoted to the superradiant phase transition and subsequent generation of entanglement between spin and momentum degrees of freedom.
In Sec.~\ref{sec:monitoring_1}, we extend the model by introducing an auxiliary cavity mode and show how it  enables continuous read-out of the system dynamics.
In Sec.~\ref{sec:PhD},  we analyze the dynamical responses and read-out strategies.  
In Sec.~\ref{sec:entanglement}, we    revisit   entanglement generation   in the presence of the auxiliary cavity mode and discuss prospects for  its non-invasive read-out. In the concluding Section~\ref{sec:concl}, we summarize our findings and discuss follow-up directions.

\section{\label{sec:Model}Model}

\begin{figure}
        \includegraphics[width=\linewidth]{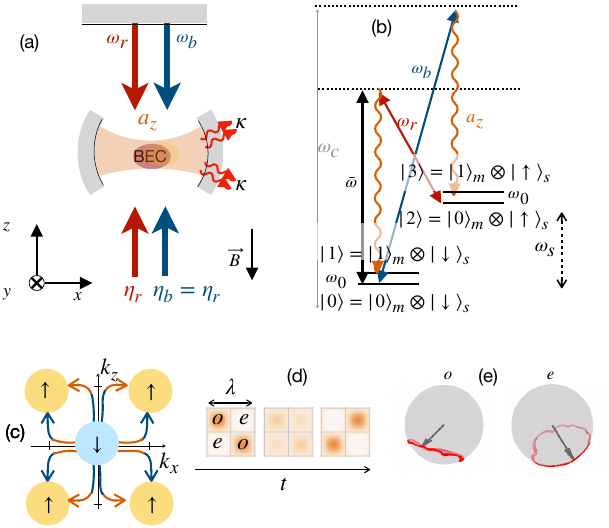}
    \caption{(a) Schematics of the experimental setup and (b) corresponding level scheme for the model~\eqref{eq:hamiltonian_b}. 
    (c)  Momentum space cartoon for  spin-flipping Raman process for a transition $\ket{0}\to\ket{3}$;  (d)  Snapshots of the real-space BEC density in the $\ket{\uparrow}_s$ manifold
    at different times;  
    (e) Dynamics in the superradiant phase of the   spin components  on even ($e$) and odd ($o$) lattice sites plotted on the Bloch spheres.   
 } 
    \label{fig:1}
\end{figure}

We consider a cavity QED configuration in which we can   address both the spin and momentum states of ultracold atoms, enabling measurement and dynamic control (see also Ref.~\cite{Ferri} for a related setup). 
Specifically, we consider Bose-Einstein condensate (BEC)  of $^{87}$Rb atoms in the $F=1$ hyperfine ground state manifold confined in a high-finesse optical cavity.
The atoms are coupled to a $z$-polarized cavity mode $a_z$
with resonance frequency $\omega_c$ and decay rate $\kappa$, extending along $x$ direction, as illustrated in Fig.~\ref{fig:1}(a).
A bias magnetic field $\vec{B}$ along the $ {z}$ direction defines the quantization axis and induces Zeeman splitting between the sub-levels of the $F=1$ hyperfine manifold. 
 We focus on two internal atomic sub-levels $\ket{m_F=1} = \ket{\downarrow}_s$ and $\ket{m_F=0} = \ket{\uparrow}_s$, and describe the condensate with
 the spinor wave function $\Psi = \left( \psi_{\uparrow}, \psi_{\downarrow} \right)^T$.

The condensate is illuminated with  transverse standing-wave laser fields
far-detuned from the electronic transitions of the atoms. These detunings allow us to effectively eliminate the contribution of excited electronic states and to focus on the near-resonant cavity-assisted two-photon transitions between an atomic momentum  state $\ket{0}_m=\ket{k_x=0,k_z=0}$ and an excited one, which reads as a coherent superposition $\ket{1}_m=\sum_{s,s'=\pm}\ket{k_x=s k,k_z=s' k}/2$.
Here, $\hbar k=2 \pi \hbar / \lambda$ indicates the recoil momentum, with $\lambda / 2=784.7 / 2 \mathrm{~nm}$ representing the period of the standing-wave potential along the drive direction.

We introduce the following definitions of relevant combinations of momentum and spin states [cf.  Fig.~\ref{fig:1}(b)]
\begin{equation}\label{eq:levels}
    \begin{aligned}
        \ket{0}&=\ket{0}_m\otimes\ket{\downarrow}_s,\\
        \ket{1}&=\ket{1}_m\otimes\ket{\downarrow}_s,\\
        \ket{2}&=\ket{0}_m\otimes\ket{\uparrow}_s,\\
        \ket{3}&=\ket{1}_m\otimes\ket{\uparrow}_s,\\
    \end{aligned}
\end{equation} 
limiting our consideration to a four-level model of the system, which will be further justified in the following. In this notation, even states $\ket{0}\textrm{ and }\ket{2}$ are momentum ground states that correspond to the homogeneous condensate density in  real space, while odd states $\ket{1}\textrm{ and }\ket{3}$ are excited momentum states and correspond to a modulation of the atomic density in  real space: $\ket{1}_m \propto \cos kx \cos kz$  (see also Ref.~\cite{baumann2010dicke}). We also introduce boson annihilation and creation operators $c_{0,\ldots,3},~ c_{0,\ldots,3}^{\dagger},$ $[c_i,c_j^{\dagger}]=\delta_{i,j},$ which  describe the annihilation and creation of a particle in these four state manifold of Eqs.~\eqref{eq:levels}.

We consider Raman processes  that simultaneously couple internal (spin) and external (momentum) atomic degrees of freedom~\cite{Ferri,LevSpinorSelfOrdering,PhysRevA.97.043858}. 
These processes are mediated by the interaction of the cavity mode $a_z$ and two classical  driving fields with coupling strength  $\eta_b=\eta_r=\eta$ and frequencies $\omega_b,$ $\omega_r,$  with $2\bar{\omega}=\omega_b+\omega_r$, $\delta=\omega_b-\omega_r$. 
In this context, the laser at frequency $\omega_b$ facilitates the transition between states $\ket{0}$ and $\ket{3}$, from the ground momentum state $|0\rangle_m$ to the excited momentum state $|1\rangle_m$, accompanied by a spin flip from $|\downarrow\rangle_s$ to $|\uparrow\rangle_s$ and vise versa. Conversely, the laser at frequency $\omega_r$ induces a similar transition, $\ket{2}\leftrightarrow\ket{1}$, accompanied by a spin flip from $|\uparrow\rangle_s$ to $|\downarrow\rangle_s$, 
cf. Fig.~\ref{fig:1}(b,c). These two cavity-assisted Raman processes  mediate an interaction of the form
$\eta \left(a_z+a_z^{\dagger}\right) \cos k x \cos k z (F^++F^-),$ where the spin operators $F^{\pm}$ couple \ox{neighboring spin levels}
[cf. momentum cartoon in Fig.~\ref{fig:1}(c) and Appendix~\ref{sec:derivation} for the model details].  After integrating over the spatial extent of the condensate, the Hamiltonian of the effective model  reads 
\begin{equation}\label{eq:hamiltonian_b}
\begin{aligned}
    H &=\omega_z a_z^{\dagger} a_z+(\omega_0-\omega_s)S_{12}^z+(\omega_0+\omega_s)S_{03}^z\\
    &+2 \eta \left(a_z+a_z^{\dagger}\right)\left(S_{12}^x+S_{03}^x\right),
    \end{aligned}
\end{equation}
where $\omega_z$ is the cavity detuning \ox{(cf. Appendix~\ref{sec:derivation})}, $\omega_0$ is the double recoil energy $\omega_{\operatorname{rec}}=k^2/(2M),$ which atoms acquire in the two-photon process, and $\omega_s$ is the effective splitting between the two spin manifolds. We set $\hbar=1$ to keep the notation compact.

The cavity boson field $a_z$ satisfies the commutation relations $[a_z,a_z]=0,$ $[a_z,a_z^{\dagger}]=1.$
The collective pseudo-spin operators $S$ are built as  projectors between  the macroscopically occupied spin-momentum levels $S_{ij}^-=\ket{i}\bra{j}=c_{i}^{\dagger}c_{j}$,  $S_{ij}^+=(S_{ij}^-)^{\dagger},$  $S_{ij}^z=(c_{j}^{\dagger}c_{j}-c_{i}^\dagger c_i)/2,$ $S_{ij}^x=(S_{ij}^-+S_{ij}^+)/2$ 
\ox{with $(i,j)\in\{(0,3),(1,2)\}$ } for $i=0,\ldots,3<j.$ Here, we already rescaled spin and photon operators via $S\to S/N,$ $a_z\to a_z/\sqrt{N}$ (and also $c_i\to c_i/\sqrt{N}$) as it is convenient for collective spin models~\cite{kirton2019introduction,Chelpanova_Intertwining_2023}, where $N$ is a number of atoms in the condensate.  \ox{Similarly, one can introduce other pseudo-spin operators $\mathcal{T}$, that are built from different spin states of the same momentum: $\mathcal{T}_{ij}^-=c_{i}^{\dagger }c_{j},\ldots$ where $(i,j)\in\{(0,2),(1,3)\}$, or when $(i,j)\in\{(0,1),(2,3)\}$ pseudo-spin operators $J_{ij}^-=c_{i}^{\dagger }c_{j}$ correspond to transitions between different momentum states of the same internal spin. 
}

The dynamics of the system are described by the Lindblad master equation for the density matrix $\rho$
\begin{equation}\label{eq:Lindblad_b}
    \begin{aligned}
& \frac{d}{d t} \rho=-i[H , \rho]+\kappa \mathcal{D}(\rho) \\
& \mathcal{D}(\rho)=2 a_z \rho a_z^{\dagger}-\left\{a_z^{\dagger} a_z, \rho\right\}.
\end{aligned}
\end{equation}
where cavity losses  account for the finite lifetime \ox{$\propto 1/\kappa$} of the cavity photon $a_z.$

Our proposal is inspired by the experiments reported in Refs.~\cite{Ferri,dogra2019dissipation} that typically involve a substantial number of atoms,  around $N \approx 10^4-10^5$, and deviations from mean-field behavior only become pronounced at extremely long timescales. The coupling between photons and atoms is collective [cf. Eq.~\eqref{eq:hamiltonian_b}], and  results in a suppression of light-matter correlations by a factor of $1/N$. 
The dynamics are thus well captured by mean-field  equations of motion, which we report in    Appendix~\ref{sec:EoM} for completeness. However, 
correlations between the spin and momentum degrees of freedom within the condensate play a significant role in entanglement dynamics, as 
we show in detail in the following Section.

\bigskip

 The dynamics in Eq.~\eqref{eq:Lindblad_b}  possess a $\mathbb{Z}_2$ symmetry characteristic of Dicke models~\cite{DickeDicke,DimerDicke,Ferri,LevSpinorSelfOrdering,NagyDickeModelPhaseTransition}: it is invariant under the transformation $ ( a_z, S^x )\leftrightarrow(- a_z ,- S^x),$ where 
 $S^x=S_{03}^x+S_{12}^x$ [cf. Eq.~\eqref{eq:hamiltonian_b}].  
 When this symmetry is spontaneously broken, the system undergoes a   phase transition.  In the thermodynamic limit, the transition can shift the system from the trivial normal state with the empty cavity mode $n_z=\langle a_z^{\dagger}a_z\rangle=|\langle a_z\rangle|^2=0$ and all spins polarized along the $z$ direction to the superradiant (SR) phase with the non-zero occupation of the cavity mode and finite $ {x}$ component of the spin, namely $n_z\ne 0$ and $\langle S^x\rangle \ne 0$ (cf. Appendix~\ref{app:dicke} or Ref.~\cite{kirton2019introduction} for more details). Throughout this paper, we employ $\langle\cdot\rangle$ to denote expectation values of observables.

  When $\omega_s=0$, the critical coupling at which the transition to the SR phase takes place read $\eta_{\omega_s=0}^c=[\omega_0(\omega_z^2+\kappa^2)/(4\omega_z)]^{1/2}$~\cite{DickeDicke,kirton2018superradiant}, while for $\omega_s\ne 0$ case, the critical coupling becomes sensitive to the initial conditions. 
 This sensitivity is rooted in the different effective level splitting for pseudo-spins $S_{03}$ and $S_{12}$, $\omega_0\pm\omega_s$. The specific distribution of particles between the two pseudo-spins, $S_{03}$ and $S_{12}$, gives rise to distinct effective level splittings between excited and ground momentum states, and thus different critical couplings.
 A similar dependence on the initial state also emerges when there is disorder in the coupling constants, as discussed in Refs.~\cite{mivehvar2023unconventional,marsh2023entanglement}.

To illustrate this dependence, consider initialization of the system in a mixture of the atoms in the ground momentum state, $|k_x,k_z\rangle=|0\rangle_m$, with two different magnetic numbers; namely, we prepare $N_0=N\langle c_0^{\dagger}c_0\rangle=\mu N$ particles in level $|0\rangle $ and $N_2=N\langle c_{2}^{\dagger}c_2\rangle=(1-\mu)N$ particles in level $|2\rangle$, where $\mu\in[0,1]$. 
This results in the critical coupling    (see Appendix~\ref{app:dicke})
\begin{equation}\label{eq:critical_coupling}
    2 \eta^c= \begin{cases}\sqrt{\frac{\left(\omega_0^2-\omega_s^2\right)\left(\omega_z^2+\kappa^2\right) / \omega_z}{\left(\omega_s+\omega_0\right)-2 \mu \omega_s}}, & \omega_s<\omega_0 \\ \sqrt{\frac{\left(\omega_s^2-\omega_0^2\right)\left(\omega_z^2+\kappa^2\right) / \omega_z}{\left(\omega_s+\omega_0\right)-2 \mu \omega_0}}, & \omega_s>\omega_0.\end{cases}
\end{equation}
If the system is prepared in the spin-polarized state, the expression for the critical coupling simplifies to $2\eta_c=[(\omega_0\pm\omega_s)(\omega_z^2+\kappa^2)/\omega_z]^{1/2}$ for $\mu=0,1$, coinciding with the critical coupling in Ref.~\cite{Mivehbar2019}.

\bigskip 

In order to study the onset of the SR phase on a microscopic level, we evaluate the condensate $|\psi_{\uparrow,\downarrow}|^2$ and the spin $\Psi^{\dagger}\sigma_i\Psi/2$ densities.  Here, we use spin-1/2 Pauli matrices $\sigma_i$ instead of spin operators $F$ of the original problem to highlight the two-level internal spin structure of the effective model. The spinor $\Psi=(\psi_{\uparrow},\psi_{\downarrow})^T$ has  components  $\psi_{\downarrow}=\langle c_0\rangle +\langle c_1\rangle \cos kx 
\cos kz$, and $\psi_{\uparrow}=\langle c_2\rangle +\langle c_3\rangle \cos kx\cos kz,$ cf. Appendix~\ref{sec:derivation}.

When $\mu\in (0,1),$ both spin and condensate density are time dependent. We show a few snapshots of the condensate density at different times in Fig.~\ref{fig:1}(d) along with the spin for even and odd sites [Fig.~\ref{fig:1}(e)); spin components are evaluated in the center of lattice cells of the size $\lambda/2\times\lambda/2$].  We explain such time-dependence from the fact that the correct order parameter that captures transition to the superradiant phase is the spin density {integrated over the space}, $\int d\mathbf{r} \Psi^{\dagger}\sigma^x\Psi\cos kx\cos kz$ (see derivation of the Hamiltonian in Appendix~\ref{sec:derivation}). 
On the other hand, the $x$ component of spatial spin profile (spin density) contains a time-independent contribution  $\propto \operatorname{Re}(\langle c_0^{\dagger} c_3\rangle+\langle c_2^{\dagger}c_1\rangle)\cos kx\cos kz^{}
\propto (\langle S_{03}^x\rangle +\langle S_{12}^x\rangle )\cos kx\cos kz^{}$, which exactly reflects spontaneous breaking of the $\mathbb{Z}_2$ symmetry, and a time-dependent contribution of the form $\propto\operatorname{Re}(\langle c_2^{\dagger} c_0\rangle +\langle c_3^{\dagger} c_1\rangle\cos^2kx\cos^2 kz)\propto(\langle \mathcal{T}_{02}^x\rangle  +\langle \mathcal{T}_{13}\rangle ^x\cos^2kx\cos^2 kz)$. Here, both $\langle \mathcal{T}^x(t)\rangle\propto \cos(\Omega t)$ [cf. expression for $\Omega$ in Appendix~\ref{app:dicke}] and are zero only if $\mu=0$ or $\mu=1$ (this particular case has been studied in~\cite{LevSpinorSelfOrdering,Ferri}). Such precession of $\langle \mathcal{T}\rangle $ originates from the fact that pseudo-spin species $ S$, $\mathcal{T}$ and $J$ are built as bilinears of the same boson operators $c_0,\ldots,c_3$. 
Similarly, the total density $|\psi_{\downarrow}|^2+|\psi_{\uparrow}|^2$ contains time-dependent contribution of the form $(\langle c_0^{\dagger}c_1\rangle+\langle c_2^{\dagger}c_3\rangle+\langle c_1^{\dagger}c_0\rangle+\langle c_3^{\dagger}c_2\rangle)\cos kx \cos kz\propto ( \langle J_{01}^x\rangle +\langle J_{23}^x\rangle)\cos kx \cos kz$ which vanishes only if $\mu=0,1.$ 
Otherwise, the spin and density distribution along the lattice are time-dependent. 
In the following section, we show the impact of such time dependence on the dynamics of entanglement  between spin and momentum degrees of freedom.

  \ox{Following experiments in Refs.~\cite{Ferri,dogra2019dissipation}, we maintain the key parameters  
$\omega_0 / (2 \pi) \approx 7.4 \mathrm{kHz}$ and $\kappa/(2 \pi)= 1.25 \mathrm{MHz}$ fixed for all simulations. The remaining detunings and coupling strengths are tunable, allowing for the exploration of a broad spectrum of dynamical regimes.}    

\begin{figure*}
    \centering \includegraphics[width=0.94\linewidth]{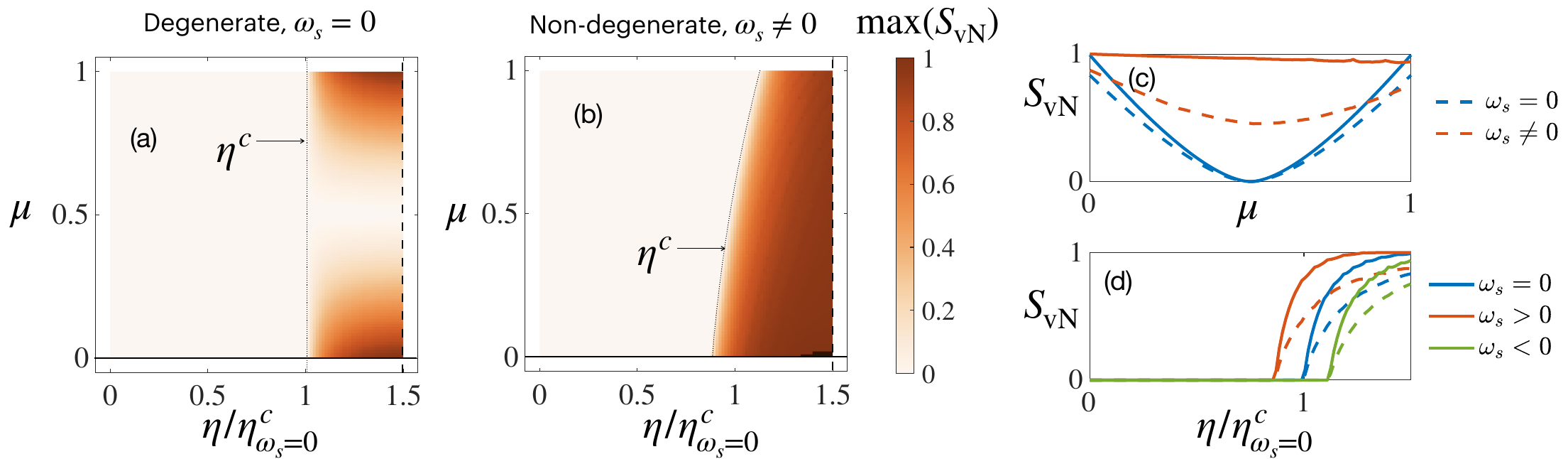}
    \caption{(a-b) Maximum of spin-momentum entanglement $S_{\operatorname{vN}}$ as a function of $\eta/\eta_{\omega_s=0}^c $ and $\mu$. Entanglement is built up as we enter the superradiant phase.  Here (a) $\omega_s=0$ and  (b) $\omega_s=\omega_0/4$. (c) Maximal (solid lines) and time-averaged (dashed lines) entanglement entropy as a function of $\mu$ for different values of $\omega_s$ and $\eta/\eta_{\omega_s=0} ^c = 1.5.$
    (d) Maximal (solid lines) and time-averaged (dashed lines)   entanglement entropy
    as a function of photon-matter coupling $\eta $ for  $\mu=0$. Different colors indicate different $\omega_s$. }
    \label{fig:entanglemet_phase_d_1}
\end{figure*}

\section{Spin-momentum entanglement and superradiant dynamics\label{sec:ent_ph_tr_1}}

Although correlations among different atoms are negligible, our platform offers a route to engineer   robust entanglement between spin and   momentum degrees of freedom within the bosonic condensate trapped in the cavity. 
For instance, assume all atoms are initially prepared in the state $|0\rangle=|0\rangle_m \otimes|\downarrow\rangle_s$. Through the interaction with the cavity mode $a_z$,  the atoms are coupled to the state $|3\rangle=|1\rangle_m \otimes|\uparrow\rangle_s$ as  $2\eta(a_z+a_z^{\dagger})S_{03}^x\ket{0}=\eta(a_z+a_z^{\dagger})\ket{3}$. Thus, in the SR phase with the non-zero cavity field ($\langle a_z\rangle \ne 0$), the cavity-mediated interaction gives rise to a non-separable spin-momentum state. 
The corresponding state of each atom reads 
\begin{equation}
  \ket{\psi}=\langle c_0\rangle |0\rangle_m \otimes|\downarrow\rangle_s+\langle c_3\rangle |1\rangle_m \otimes|\uparrow\rangle_s, 
\end{equation}
with $\langle c_0\rangle\ne 0$ and $\langle c_3\rangle\ne 0,$ and 
 $|\psi\rangle$ is   a non-separable entangled state of spin and momentum.  Our results will revolve around the dynamical manipulation of this form of entanglement.  

In order to quantify spin-momentum entanglement, we use the von Neumann entropy 
\begin{equation}\label{eq:vonNeumannEntropy}
    S_{\operatorname{vN}}=-\Tr\left(  {\tilde{\rho}}\log_2 {\tilde{\rho}}\right)
\end{equation}
\noindent with $\tilde{\rho}$   the reduced density matrix after tracing out spin or momentum states, cf. Appendix~\ref{sec:Entanglement}. When the system is in a product state of spin and momentum, the entanglement vanishes and $S_{\operatorname{vN}}=0$. With the definition in Eq.~\eqref{eq:vonNeumannEntropy}, a maximally entangled state has $ S_{\operatorname{vN}}=1.$ We   also compute negativity~\cite{negativity} and concurrence~\cite{wootters2001entanglement,Zou_2022},
which are  more reliable witnesses of   entanglement in open   systems~\cite{PhysRevA.65.032314}.
However, they show the same qualitative behavior as $S_{\operatorname{vN}}$ (cf. Appendix~\ref{sec:Entanglement}), and thus we restrict our analysis to the von Neumann entropy for its simplicity.

 By adjusting coupling $\eta$ and the initial state of atoms, we compute a dynamical phase diagram, which captures maximal $S_{\operatorname{vN}}$ reached during evolution, see Fig.~\ref{fig:entanglemet_phase_d_1}. The system is initially prepared in the normal state with $N_0=\mu N$ atoms in the state $\ket{0}$, after which we rapidly increase the coupling $\eta$ to a specified value. The equations of motion describing this process can be found in Appendices~\ref{sec:EoM} and \ref{sec:Redfield}. Note that the collective description of the model, adapted in this work, is valid only when the system is prepared in the permutation invariant state; otherwise, dynamics become more complicated as discussed in~\cite{PhysRevB.101.214302,PhysRevB.108.054301,iemini2023dynamics,Riccardo}. Below, we analyze the entanglement properties for both the degenerate case $\left(\omega_s=0\right)$ and the non-degenerate case $\left(\omega_s \neq 0\right)$, showcasing the potential for achieving either a stationary or a time oscillating amount of entanglement, respectively. 
 
 \subsubsection{Degenerate $\omega_s=0$ case}
 
The  amount of $S_{\operatorname{vN}}$ reached during dynamics in the degenerate case is shown in Fig.~\ref{fig:entanglemet_phase_d_1}(a).
Depending on the initial configuration, the maximum amount of entanglement in the system can vary from $\max( S_{\operatorname{vN}})=0$ to $\max( S_{\operatorname{vN}})=1.$ Specifically, when $\mu=0.5,$ the momentum configuration of each spin component reads exactly the same, and the state becomes separable. Conversely, when $\mu=0$ or $1,$ one can reach a maximally entangled state; the dependence of the entanglement in the system as a function of $\mu$ is shown with blue lines in Fig.~\ref{fig:entanglemet_phase_d_1}(c). 
On the other hand, the amount of entanglement in the SR phase depends on the coupling $\eta$, [cf. dependence of the entanglement entropy as a function of coupling for $\mu=0$  in panel (d), blue lines]. Here, entanglement increases with the coupling which can be qualitatively understood as follows. When $\eta \approx\eta ^c,$ almost all atoms occupy the ground momentum state and $\ket{\psi}\propto \ket{2}$, which is separable in terms of spin and momentum. However, as we increase coupling, the population of the excited momentum level $\ket{1}$ increases and the spin-momentum state of the system becomes non-separable, approaching a maximally entangled state deep in the SR phase. {Importantly, by solving the dynamics of the system without cutting off the higher momentum state, we check that for large couplings, most of the atoms occupy momentum states $\ket{0}_m,~\ket{1}_m$. The states $\ket{0,2k}$ and $\ket{2k,0}$ are significantly less populated during the dynamics, and we can neglect them.  
The corresponding equations of motion are given in Appendix~\ref{sec:EoM}. 
}

\begin{figure*}
    \centering   \includegraphics[width=0.8\linewidth]{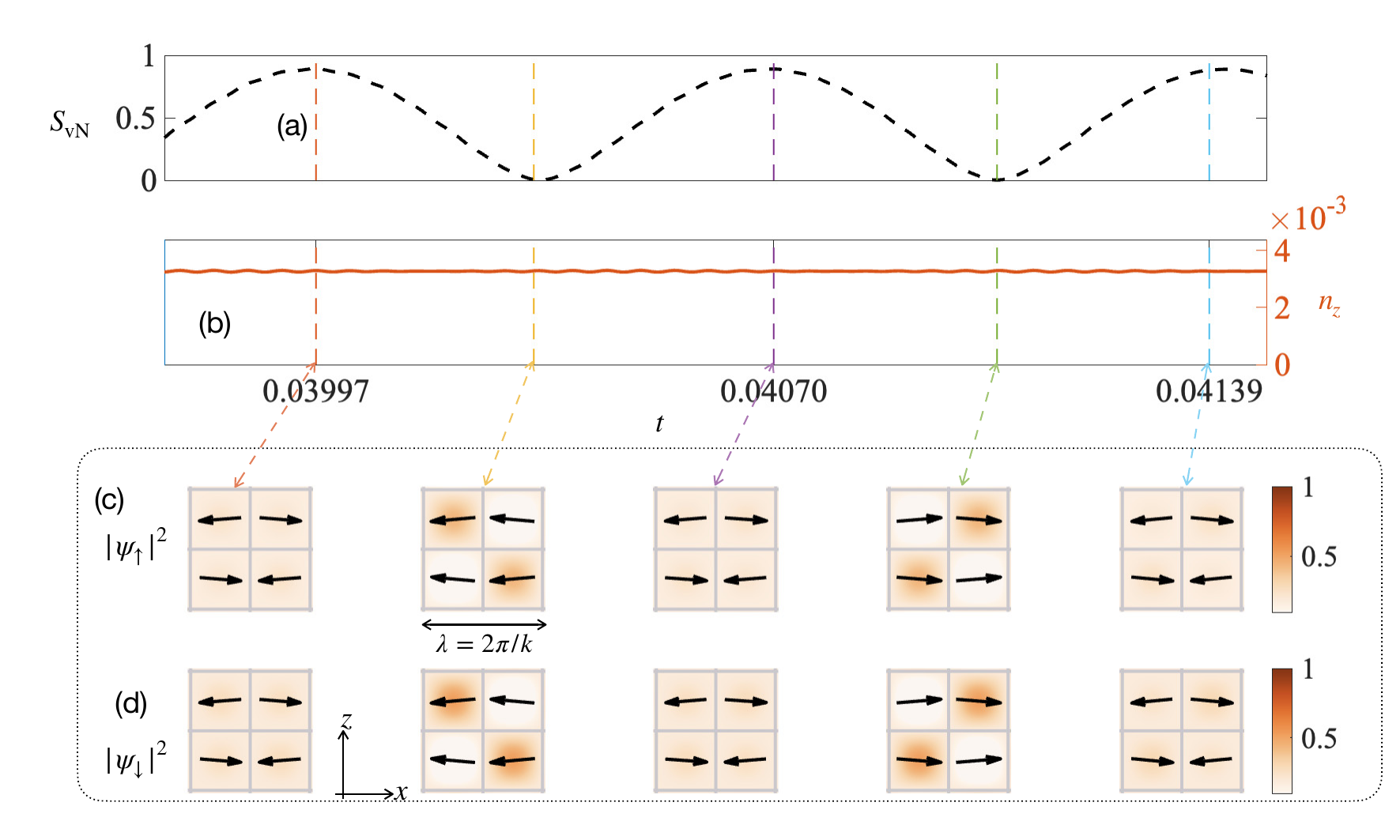}
    \caption{(a) Dynamics of entanglement, (b) occupation of the cavity mode $a_z,$ $n_z=\langle a_z^{\dagger}a_z\rangle,$  density of the (c) upper $|\psi_{\uparrow}|^2$ and (d) lower $|\psi_{\downarrow}|^2$ spinor components. Black arrows in panels (c-d) indicate local spin texture $\Psi^{\dagger} \sigma_i \Psi/2$, evaluated in the centers of each cell and projected onto the $xz$ plane. Dashed lines indicate time moments when the entropy takes extremum values and is added to guide the eye. Entropy is maximized when the state $\ket{\psi}$ is a non-separable combination of two spin-momentum states, and   SR manifests as a spin checkerboard lattice, i.e., projection of spin onto $x$ axis takes opposite signs on even and odd lattice cells. When entanglement vanishes, both spinor components have similar density profiles and the sign of projection of the spin magnetization onto the $x$ axis.
    Dynamics are simulated for $\mu=0.5,$ $\eta =1.6\eta ^c,$ $\omega_s=\omega_0/4,$ $\omega_z=\kappa.$} 
    \label{fig:entanglement_lattice}
\end{figure*}

 \subsubsection{Non-degenerate $\omega_s\ne 0$ case}

When $\omega_s\ne 0,$ the dynamical behavior of the entanglement entropy changes compared to the degenerate case [see  Fig.~\ref{fig:entanglemet_phase_d_1}(b)].  
First, the critical coupling depends on $\mu$, cf. Eq.~\eqref{eq:critical_coupling}.
Second, when $\mu=0.5,$ entanglement decreases but does not disappear completely [red lines in panel (c)]. Here, one can notice the difference between the  maximal value of the entanglement (dashed line)  and the period-averaged value (solid line), indicating an oscillatory behavior in time.
The exceptional case is $\mu=0,1$ where the system evolves towards a stationary SR state. From the standpoint of spin-momentum correlations, this state has maximum (non-oscillatory) entanglement when compared with occurrences at other values of $\mu$.

Fig.~\ref{fig:entanglement_lattice} depicts the dynamics of the entanglement entropy in the non-degenerate case after the quench during the period (a), along with the (b) occupation of the cavity field $n_z$ and (c-d) spin (arrows) and density distributions of atoms. 
In panels (c-d) with arrows, we plot the projection of this spin density on the $xz$ plane, evaluated in the center of each lattice cell of size $\lambda/2\times\lambda/2$, which is formed by the interference of the laser and cavity fields.  

In Fig.~\ref{fig:entanglement_lattice}, maximally entangled configurations correspond to the case when the checkerboard lattice is formed by the spin degree of freedom (arrows), while the condensate density is periodically modulated with the period $\lambda/2.$ On the contrary, in the configuration with vanishing entanglement,  the sign of the spin projection $\Psi\sigma^x\Psi$  is fixed for all lattice cells, and the condensate density modulation occurs with a period $\lambda.$ 
Such dynamics persist in time without any sign of relaxation [see Appendices~\ref{sec:EoM} and~\ref{app:dicke} for more details].

Such dynamical behavior emerges from the time-dependent components of spin and condensate densities, if the system is initialized in the mixed internal spin state. 
While the total atom number $\int d\mathbf{r}|\Psi|^2=N$, and and transverse magnetization  $\int d\mathbf{r}\Psi^{\dagger}\sigma^x\Psi/N\propto \langle S^x\rangle$, are conserved quantities in the steady state, their local distributions can exhibit exhibit complicated time-dependence, driven by the precession of pseudo-spins $\langle J\rangle$ and $\langle \mathcal{T}\rangle $:
When $\langle J^x\rangle$ and $\langle \mathcal{T}^x\rangle$ reach zero, the local distribution of spin and condensate densities are identical to the one obtained starting from the polarized spin state, and the entanglement entropy becomes maximal. When $\langle J^x\rangle$ and $\langle \mathcal{T}^x\rangle $ deviate from zero, the distribution of atoms among space and spin levels changes, decreasing the entanglement entropy. 
{Thus, the time dependence of spin and momentum states results in oscillations of the entanglement entropy.} 
This is one of the striking features of our quench protocol: we can steer entanglement dynamics toward an oscillatory regime that persists up to the operational timescales of the experiment. 


\bigskip

Notice that entanglement is generated during dynamics starting from product spin-momentum states, and thus, the cavity photon has an active role in building spin-momentum correlations via light-induced interactions. 
The non-zero cavity field $\langle a_z\rangle$ in the SR phase mediates an effective interaction among atoms, which is responsible for entangling them in a non-separable spin-momentum state. 
At the same time, the role of cavity losses is essential. They steer dynamics towards the fixed point of the Lindbladian, with the remarkable consequence that all entanglement properties derived in the presence of photon losses are robust if compared with what would be achieved with a coherent drive~\cite{galitski2013spin,lorenzo2017quantum,kale2020spin}. For instance, by replacing the cavity field with a time-dependent drive in $H$ [Eq.~\eqref{eq:hamiltonian_b}], one could also entangle spin and momentum degrees of freedom of the condensate's atoms. However, the amount of final entanglement produced would depend on details of 
the driving protocol,
such as its duration, frequency decomposition, and other specifics. More importantly, such entanglement would be highly sensitive to noise and imperfections in the drive realization~\cite{szigeti2021improving,colombo2022entanglement,greve2022entanglement}. In contrast, cavity losses induce relaxation of  atomic entanglement towards a steady state that remains resilient even for 
moderate imperfections in the initial state preparation or in the parameters set to drive dynamics into superradiance. In other words, there exists a broad basin of attraction towards   prescribed values of entanglement   given the system's initial conditions parameters.

\section{Two cavity fields setup \\ and probes of   momentum states}\label{sec:monitoring_1}

In the previous section, we have shown that cavity dissipation can be utilized to prepare the system in a steady state with desired entanglement properties. In the following sections, we show how, by using the auxiliary polarization mode $a_y$ of the cavity field, one can get access to the collective momentum state of the system in a non-destructive fashion.

\begin{figure}
    \centering
        \includegraphics[width=\linewidth]{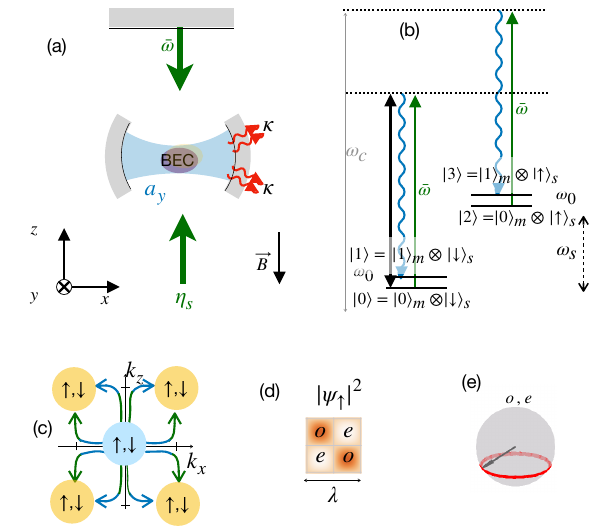}

    \caption{(a) Schematics of the experimental setup and (b) corresponding level scheme for the model~\eqref{eq:hamiltonian_s}. 
    (c) Momentum space cartoon for the spin-preserving Bragg process;  (d) steady-state density of the condensate in the SR phase;   (e) internal spin precession described by the Hamiltonian $H_s.$ Here, spontaneous symmetry breaking in $H_s$ results in a density checkerboard modulation.  
 } 
    \label{fig:level_schemes_Hs}
\end{figure}

Inspired by the experimental demonstrations in Refs.~\cite{baumann2010dicke,dogra2019dissipation} we consider a driving scheme that enables effective coupling of ground and excited atomic momentum states.
We consider a cavity-assisted Bragg  process involving the  transverse  driving field  with  amplitude $\eta_s$  and  frequency $\bar{\omega}$ 
and the cavity mode $a_y$ with detuning $\omega_y,$  decay rate $\kappa$, and linear polarization along $y$  [see Fig.~\ref{fig:level_schemes_Hs}(a,b)].  This process is reflected in the atom-cavity interaction term, $\propto \eta_s(a_y+a_y^{\dagger })\cos kx \cos k z$ (cf. Appendix~\ref{sec:derivation}). 
In this two-photon process, atoms initialized in the ground momentum state $\ket{0}_m$ can be excited to the  momentum state $\ket{1}_m$, while the internal spin state ($\ket{\downarrow}_s$ or $\ket{\uparrow}_s$) remains unchanged [cf. Fig.~\ref{fig:level_schemes_Hs}(c)]. 
The schematics  of this process are   encoded in   the Dicke Hamiltonian (see Appendix~\ref{sec:derivation})
\begin{equation}\label{eq:hamiltonian_s}
    H_s=\omega_y a_y^{\dagger} a_y+\omega_0\left(J_{01}^z+J_{23}^z\right)+2 \eta_s\left(a_y+a_y^{\dagger}\right)\left(J_{01}^x+J_{23}^x\right).
\end{equation}

\bigskip

Depending on the coupling $\eta_s,$ the system undergoes a phase transition associated with the spontaneous breaking of the $\mathbb{Z}_2$ symmetry of the Hamiltonian $H_s$, such that the Hamiltonian is invariant under the transformation $(a_y,J^x)\leftrightarrow( -a_y,-J^x)$.  When the coupling is below the critical value $\eta_s<\eta_s^c={[\left(\omega_y^2+\kappa^2\right) \omega_0^2 /\left(4 \omega_y\right)]^{1/2}}$,
the system is in the normal phase where only ground momentum states are occupied, and, respectively, $\langle J_{01}^z
\rangle+\langle J_{23}^z
\rangle=-1/2,$ $\langle J_{01}^x\rangle =\langle J_{23}^x\rangle =0,$ and the cavity is empty, $n_y=\langle a_y^{\dagger}a_y
\rangle=|\langle a_y\rangle|^2=0$ (see Appendix~\ref{app:dicke} for more details). In this phase, the condensate is homogeneously distributed within the trap without a checkerboard-like density modulation.
When  $\eta_s>\eta_s^c$, the system enters a $\mathbb{Z}_2$ symmetry-broken superradiant phase with $\langle J^x \rangle \ne 0$ and $n_y\ne 0$. 

On a microscopic  level, in the SR phase, the standing-wave driving field and the cavity field form an interference lattice potential $V\propto \cos kx \cos kz$, and the condensate density is modulated, forming the checkerboard lattice with the period $\lambda=2\pi/k,$ see Fig.~\ref{fig:level_schemes_Hs}(d). 
The density modulation originates from the condensate wave function in each spinor component.
At the same time, as it is shown in Fig.~\ref{fig:level_schemes_Hs}(e), the internal spin $\Psi^{\dagger}\sigma\Psi$  also precesses according to the model~\eqref{eq:hamiltonian_s} with the   amplitude $\propto\sqrt{\mu(1-\mu)}$ and   frequency $\propto 2\omega_s$ (see Appendix~\ref{app:dicke}).  

The interaction term in Eq.~\eqref{eq:hamiltonian_s} couples different atomic momentum states within the same spin manifold and does not generate entanglement between spin and momentum. Precisely, in the SR phase, the momentum configuration reads exactly the same for each spin manifold $\ket{\uparrow}_s$ and $\ket{\downarrow}_s$, and the state is separable. However, as we show below, competition between $H$ and $H_s$ results in a rich manifold of  dynamical responses, which can be probed in a non-destructive way by analyzing the light that leaks out of the cavity~\cite{heterodyne}. As we report further in the text, by tuning $\eta$ and $\eta_s$, it is possible to monitor the spin-momentum entanglement generated by $\eta$ in a non-invasive manner. This means that such monitoring can be achieved without substantially altering the underlying dynamics of entanglement entropy.

\subsection{Intertwined spin and momentum dynamics}

The   Hamiltonian describing the interaction of the two  cavity modes with different polarizations and four spin-momentum levels reads 
\begin{equation}\label{eq:model}
H_{\operatorname{tot}}= H+H_s -\omega_0\left(J_{01}^z+J_{23}^z\right),
\end{equation}
where the term $H_s$ [Eq.~\eqref{eq:hamiltonian_s}]   describes transitions in the momentum degrees of freedom, while   $H $   [Eq.~\eqref{eq:hamiltonian_b}] describes transitions simultaneously in momentum and spin degrees of freedom. Here, we subtract the $\omega_0 J^z$ term since it is already included in both $H $ and $H_s$, see derivation in Appendix~\ref{sec:derivation}.
 The overall dynamics of this open system are governed by the Lindblad master equation
\begin{equation}\label{eq:Lindblad}
\begin{aligned}
&\frac{d}{d t} \rho=-i[ {H_{\operatorname{tot}}}, \rho]+\kappa\mathcal{D}(\rho),\\
&\mathcal{D}(\rho)=2  {a}_y \rho  {a}_y^{\dagger}-\left\{ {a}_y^{\dagger}  {a}_y, \rho\right\}+2  {a}_z\rho  {a}_z^{\dagger}-\left\{ {a}_z^{\dagger}  {a}_z, \rho\right\}
\end{aligned}
\end{equation}
where we have also included a   finite lifetime \ox{$\propto 1/\kappa$} for the cavity photons. 

The key feature determining dynamics in this model is that the pseudo-spins   $J$ and $S$ are built as bilinears of  bosonic operators of  the same Hilbert space and, therefore, in general,  do  not commute with each other. 
One can define the     matrix
\begin{equation}\label{eq:sigma}
\Sigma_{i j}=c_i^{+} c_j=
\begin{pmatrix}
\Sigma_{00} & J_{01}^{-} & \mathcal{T}_{02}^{-} & S_{03}^{-} \\
J_{01}^{+} & \Sigma_{11} & S_{12}^{-} & \mathcal{T}_{13}^{-} \\
\mathcal{T}_{02}^{+} & S_{12}^{+} & \Sigma_{22} & J_{23}^{-} \\
S_{03}^{+} & \mathcal{T}_{13}^{+} & J_{23}^{+} & \Sigma_{33}
\end{pmatrix}
\end{equation}
which contains all the possible spin raising and lowering operators coupling   the four levels of our scheme. 
In $\Sigma_{ij}$ the diagonal elements account for the  occupation of the different atomic levels; the pseudo-spins $J$ describe transitions between different momentum states within the same spin state; the pseudo-spins $S$ describe transitions between different momentum states within neighboring spin levels, and  finally, the pseudo-spins $\mathcal{T}$ describe transition between different spin levels but with same momentum quantum number.
These operators obey a SU(4) algebra with the commutation   relations  
\begin{equation}   \left[\Sigma_{nm},\Sigma_{kl}\right]=\Sigma_{nl}\delta_{m,k}-\Sigma_{km}\delta_{n,l}.
\end{equation}
The non-commutativity of different pseudo-spin species (and thus also $[H_s,H ]\ne 0$) leads to rich dynamics~\cite{Riccardo,iemini2023dynamics}. In particular,    symmetry breaking in the subsystem  governed by $H$ can induce explicit symmetry breaking  in the Hamiltonian $H_{s}$, and vice versa, see Appendix~\ref{sec:SSB}.  
For instance,   the superradiant phase of  Hamiltonian $H$ [Eq.~\eqref{eq:hamiltonian_b}] corresponds to the spontaneous breaking of the $\mathbb{Z}_2$ symmetry of the system, such that two alternating non-zero solutions appear with $( a_z, S^x )\leftrightarrow(- a_z,- S^x).$
In terms of the underlying bosonic operators, the symmetry implies 
\begin{equation}\label{eq:transformation_main}
    \begin{aligned}
         c_{n}  & \to c_{n}e^{-i\phi_{n}}\\
a_{z} & \to a_{z}e^{i\pi}.
    \end{aligned}
\end{equation}
The requirement $S^x\to-S^x$ sets two constraints for four phases of the atomic fields, namely (cf. also Appendix~\ref{sec:SSB})
\begin{equation}\label{eq:constrains}
    \begin{aligned}
        \phi_{0}-\phi_{3} & =\pi\pm2\pi n\\
\phi_{2}-\phi_{1} & =\pi\pm 2\pi m.
    \end{aligned}
\end{equation}
 As a result, if the coupling $\eta_{s}$ is non-vanishing, the symmetry of the interacting term in the Hamiltonian $H_s $ will be explicitly broken by the emergent phase $\pm(\phi_1-\phi_3)$, $2J^x\to -(J_{01}^- e^{i(\phi_3-\phi_1)}+J_{23}^-e^{-i(\phi_3-\phi_1)}+h.c.),$ which can not be compensated by the phase of $a_y$.
This explicit symmetry breaking manifests in the onset of long-lived non-stationary dynamical responses, even though the Hamiltonian possesses a $\mathbb{Z}_2$ symmetry, and thus, it would be in general expected to relax into a time-independent steady state~\cite{PRADickeKeeling}.   
On the contrary, in the normal state, the emergent phase $\phi_1-\phi_3$ can be immediately set to $2 \pi n$ (both excited momentum states are unpopulated and $\left\langle c_1\right\rangle=\left\langle c_3\right\rangle=0$) and the spontaneous symmetry breaking does not bring any observable effect to the dynamics.
Finally,   spontaneous symmetry breaking in $H_s$ induces explicit symmetry breaking in $H.$ For a detailed discussion, refer to Appendix~\ref{sec:SSB}.

We want to emphasize that considering a four-level model is essential for obtaining the above-mentioned non-stationary phases. For instance, omitting the atomic level $\ket{0}$ in $H_{\operatorname{tot}}$ to get an effective three-level description relaxes the constraints of Eq.~\eqref{eq:constrains} and prevents dynamics arising from explicit symmetry breaking (see Appendix~\ref{sec:three_levels} for a comprehensive discussion).

\begin{figure*}
    \centering
\includegraphics[width=0.8\linewidth]{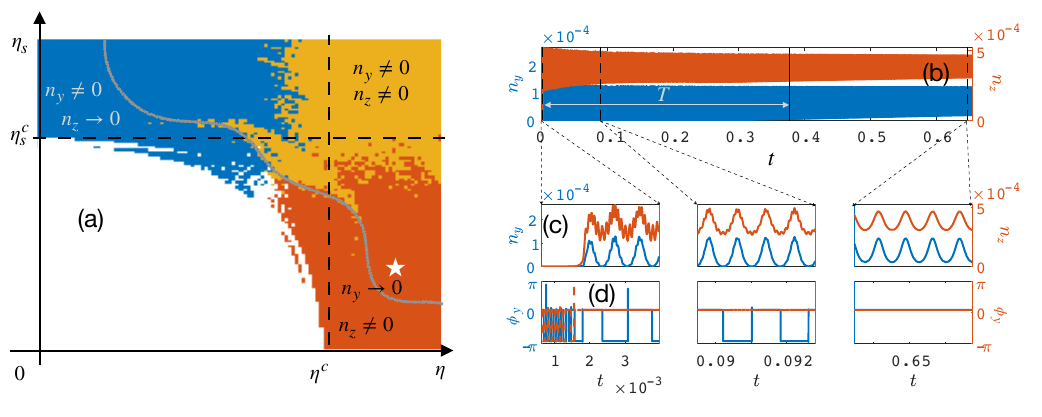}
    \caption{(a) Dynamical phase diagram with three possible self-organized phases in blue, yellow, and orange. Different phases are distinguished from the dynamics of the two cavity fields $\langle a_y\rangle $ and $\langle a_z\rangle.$ \ox{Gray line separates parameters for which the explicitly broken symmetry restores during finite (above the line) or infinite (below the line) time.}
    (b) Time evolution  of the occupation number for the two cavity modes, for parameters marked with a star in panel (a). (c) Zoom  on the dynamics of the photon number and (d) phase at the beginning, middle, and end of time evolution in panel (b). Time is given in seconds, see main text. Here we fix $\omega_y=\omega_z=5\kappa,$ $\omega_s=\omega_0/4,$ $\mu=0.75.$     
    }
    \label{fig:phase_diagram}
\end{figure*}

\section{\label{sec:PhD}Probing dynamics with the two cavity fields} 

We now discuss  the different dynamical regimes arising from the interplay of $H_s$ and $H$. We show how the auxiliary cavity field dynamics are directly linked with spin precession and entanglement entropy oscillations, facilitating continuous monitoring of the system's dynamics.

\subsection{Dynamical phase diagram}

By tuning the two couplings, $\eta$ (spin-momentum) and $\eta_s $ (momentum) below and above criticality, we generate the complete diagram of dynamical responses, reported in Fig.~\ref{fig:phase_diagram}(a). We initialize the system in the normal state~\footnote{We also tried sampling various initial states and comparing dynamics. Our general observation is the oscillatory behavior is quite generic and repeats from simulation to simulation. However, the frequency of oscillations may depend on the initial state.} and then fast ramp it at certain values of the couplings [cf. Appendix~\ref{sec:Redfield}].    

As  order parameters, we consider the mean field expectation values of two cavity fields, $\langle a_y\rangle $ and $\langle a_z\rangle.$ The choice  is convenient for two reasons. Firstly, typical experiments operate in a regime where cavity detunings, $\omega_{y,z},$ and decay rates, $\kappa$, are a few orders of magnitude larger than atomic energy scales~\cite{baumann2010dicke,Ferri,dogra2019dissipation,mivehvar2021cavity}. Consequently,
 one can adiabatically eliminate the cavity modes since on  timescales $\propto 1/\kappa$ they approach the steady state values $\langle a_y\rangle \approx -2\eta_s (\langle J_{01}^x\rangle +\langle J_{23}^x\rangle )/(\omega_y-i\kappa)$, $\langle a_z\rangle \approx -2\eta  (\langle S_{03}^x\rangle +\langle S_{12}^x\rangle)/(\omega_z-i\kappa)$~\cite{atom_only}. Thus, the cavity fields $\langle a_y\rangle$ and $\langle a_z\rangle$ offer direct information about  momentum $\langle J^x\rangle $ and spin-momentum $\langle S^x\rangle$ coherences in the system. 
 Notice that the na\"ive elimination of the cavity field at the level of the generator of dynamics would result in a lack of  relaxation, which is an artifact (in the Dicke model, the decay appears at the higher order of perturbation theory; see Refs.\cite{Jager,Ferri,PRLKeeling} for a comprehensive discussion). Indeed, in order to extract the dynamical responses in Fig.~\ref{fig:phase_diagram}, we adopt a Redfield master equation approach~\cite{atom_only,Jager}; the corresponding equations of motion are reported in  Appendices~\ref{sec:EoM},~\ref{sec:Redfield}.
 
The second reason to use cavity fields as  order parameters is their experimental accessibility. Using heterodyne detection~\cite{heterodyne}, which gives access to the magnitude and phase of the cavity fields,  it is possible to conduct continuous non-destructive measurements of the system. 
In contrast, imaging the condensate's spin and density distribution constitutes a destructive measurement, requiring numerous experiment repetitions to reconstruct  dynamics.

The system exhibits a variety of  self-organization transitions, distinguishable  by the dynamics of the cavity fields $\langle a_{y}\rangle$ and $\langle a_{z}\rangle.$
Firstly, when both couplings are smaller than the critical ones [white region in Fig.~\ref{fig:phase_diagram}(a)], the system remains in the normal state with zero occupation of the cavity fields.  In terms of atomic degrees of freedom, the internal atomic pseudo-spin precesses with the frequency $2\omega_s$ and amplitude given by $\sqrt{1/4-\langle \mathcal{T}^{z}\rangle^{ 2}}=\sqrt{\mu(1-\mu)}$.

By increasing $\eta$ above the critical value and keeping $\eta_s <\eta_s ^c$, the system undergoes a phase transition to the SR phase, associated with breaking of the $\mathbb{Z}_2$ symmetry of $H$ [cf. Eq.~\eqref{eq:hamiltonian_b}].  The occupation of the cavity mode $\langle a_z\rangle 
,$ together with the pseudo-spin $\langle S^x\rangle $ become non-zero [see red region in Fig.~\ref{fig:phase_diagram}(a)]. On the other hand, according to the  transformation~\eqref{eq:transformation_main}, this spontaneous symmetry breaking also induces explicit symmetry breaking  in $H_s$ [Eq.~\eqref{eq:hamiltonian_s}], namely, the interaction term gains a phase $\pm(\phi_1-\phi_3).$ As a consequence, the pseudo-spin $\langle J^x\rangle $ starts precessing with a zero time average,  resulting in periodic development of $\langle a_y\rangle \propto \langle J^x\rangle.$  In this way, subsystem~\eqref{eq:hamiltonian_s} experiences superradiance from the interaction with the subsystem~\eqref{eq:hamiltonian_b}; otherwise, since $\eta_s <\eta_s ^c$, the pseudo-spin $\langle J^x\rangle $ together with the cavity field $\langle a_y\rangle$ remain in the normal state.

In the experiment, this dynamical phase can be discerned by measuring both the photon number and phase of the two cavity fields, as illustrated in Fig.~\ref{fig:phase_diagram}(b-d). Following the fast ramp at $t=0$, the observable $n_z=\langle a_z^{\dagger} a_z\rangle $  approaches a non-zero value. Simultaneously, the photon number of the second mode, denoted as $n_y=\langle a_y^{\dagger} a_y\rangle $, undergoes oscillations, transitioning from zero to a finite value. At each instance when $n_y$ returns to zero, the phase $\phi_y=\arg \left(\langle a_y\rangle \right)$ experiences a discrete shift of $\pi$, signifying that $\langle a_y\rangle $ undergoes a sign reversal, as shown in Fig.~\ref{fig:phase_diagram}(c,d). Note that depending on parameters, such a regime with the zero-averaged $\langle a_y\rangle$ and periodic jumps of $\phi_y$ can take finite time, which we denote $T$ in Fig.~\ref{fig:phase_diagram}(b). \ox{This time is linked to the propensity of the system to restore explicitly broken symmetry and, as we show in the following section, the finiteness of $T$ can be related to the possibility of non-invasive continuous monitoring of the entanglement dynamics. In Fig.~\ref{fig:phase_diagram}(a) for parameters below the gray line in panel (a) $T\to\infty$, while for parameters above the line, it takes a finite value.}

In the regime where the field $\langle a_y\rangle $ oscillates with zero average, the precession frequency of the pseudo-spin $\langle J^x\rangle $ can be calculated as the inverse of the time interval over which the phase of $\langle a_y\rangle$ changes by $2 \pi$. Simultaneously, the amplitude of $\langle J^x\rangle $ oscillations can be deduced from the maximum value of $n_y$ during one period, $\operatorname{ampl}(\langle J^x\rangle )=[\max(n_y)(\omega_y^2+\kappa^2)]^{1/2}/(2\eta_s)$.

The time evolution shown in Fig.~\ref{fig:phase_diagram}(b-d) is not unique but depends on the phase that is initially imprinted in the boson $c_i$ (pseudo-spins $\Sigma_{ij}$) operators.
Different initial conditions can lead to dephasing and variations in the amplitudes and frequencies for different observables due to the nonlinear nature of the problem. 
However, as we have checked numerically, the oscillatory behavior in Fig.~\ref{fig:phase_diagram}(c-d) is generic for different realizations of the initial conditions, meaning one can observe oscillations of the magnitude of the cavity fields and also periodic jumps of the phase of the auxiliary cavity field.

\begin{figure*}
\includegraphics[width=0.8\linewidth]{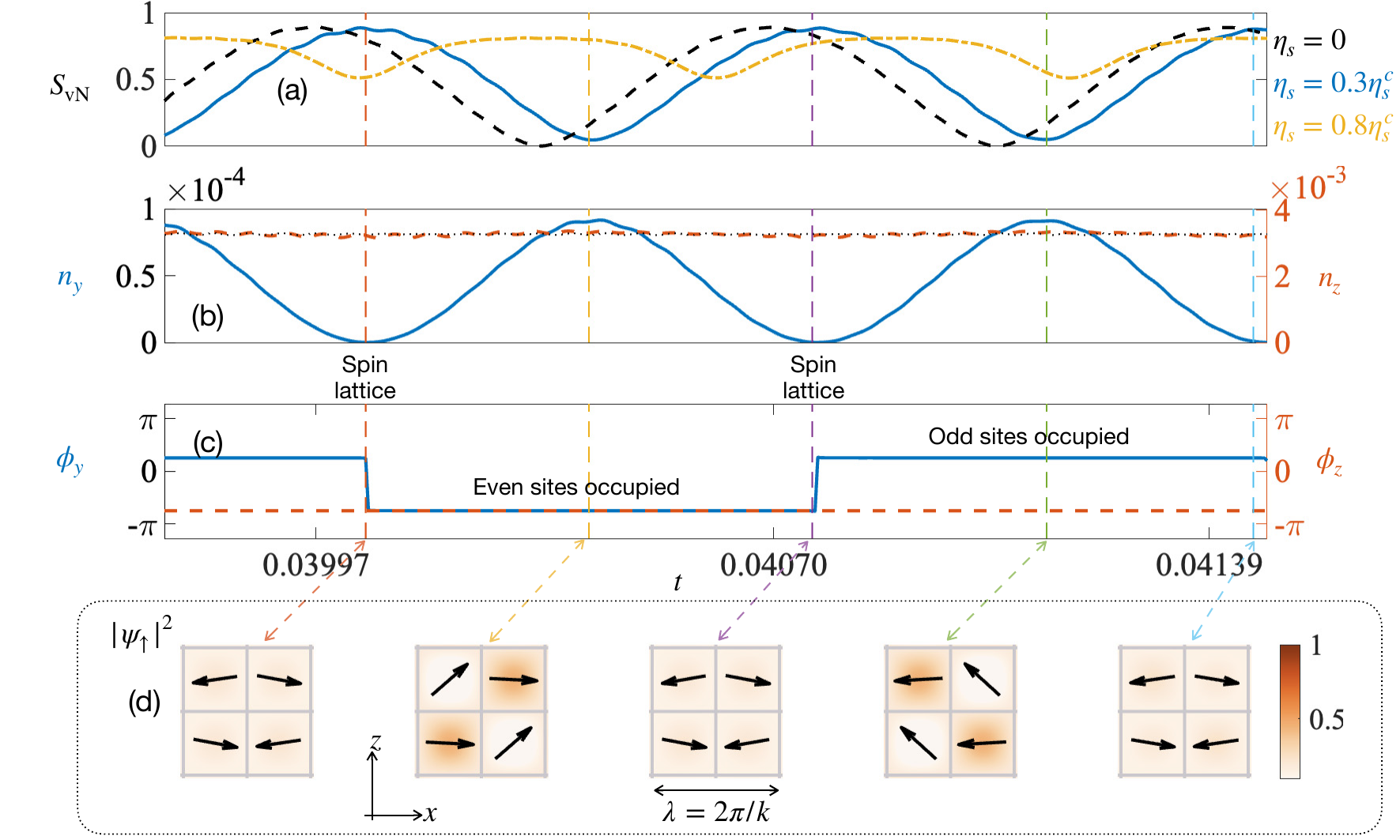}
   \caption{(a) Dynamics of the entanglement entropy, (b) occupation of the cavity modes, (c) phases of the cavity modes, and (d) real space density in the red region in the phase diagram. The black line in panel (a) corresponds to the $\eta_s=0$ case, while red and blue lines correspond to the dynamics with non-zero $\eta_s\ll \eta_s^c.$ From oscillations of the cavity field phase $\phi_y,$ one can recover the structure of the real-space checkerboard lattice: when 
   $\phi_y> 0$
   odd sites are occupied, and when 
   $\phi_y<0$
   even sited are more occupied. The $ n_y=0$ case corresponds to the maximally entangled state in which the checkerboard is formed by the projection of the spin on the $x$ axis rather than the density lattice. 
    }
   \label{fig:lattice_photon}
\end{figure*}

 In the opposite limit, 
when $\eta_s>\eta_s^c$  and $\eta $ is below its critical value [blue region in Fig.~\ref{fig:phase_diagram}(a)], the  transition to the SR phase takes place in cavity field $\langle a_y\rangle $ and pseudo-spin $\langle J^x\rangle $, while cavity mode $\langle a_z\rangle$ experiences oscillations with  zero time-average. 
These oscillations  appear due to the explicit symmetry breaking in Hamiltonian $H$ in Eq.~\eqref{eq:hamiltonian_b} and subsequent precession of the pseudo-spin $\langle S^x \rangle.$  Similarly to the previous case, the precession period is equal to the time interval during which  $\phi_z=\arg(\langle a_z\rangle )$ changes by $2\pi,$ and the amplitude of $\langle S^x\rangle $ oscillations is $\operatorname{ampl}(\langle S^x\rangle )=[\max(n_z)(\omega_z^2+\kappa^2)]^{1/2}/(2\eta)$.

Finally, when both couplings are above the critical ones [see the yellow region in Fig.~\ref{fig:phase_diagram}(a)], both cavity modes, $\langle a_y\rangle $ and $\langle a_z\rangle,$ become non-zero, and the symmetry of both $H_s$ in Eq.~\eqref{eq:hamiltonian_s} and $H$ in Eq.~\eqref{eq:hamiltonian_b}
 are spontaneously broken in a self-consistent way. Here, both cavity fields have fixed phases, while their magnitudes can oscillate while the system approaches  a (stationary) steady state.

\subsection{Slow relaxation in multi-level Dicke model}

The oscillations shown in Fig.~\ref{fig:phase_diagram}(b) persist 
 far longer than the operational timescales of the experiment. Below, we discuss the mechanism that induces such prolonged relaxation in the dissipative model~\eqref{eq:model}.

  The evolution  of   energy in the two-level Dicke model during   relaxation is given by 
$\dd \langle E\rangle/\dd t=\kappa \mathcal{D}(\omega_c a^{\dagger}a+2\eta (a+a^{\dagger})J^x)$, which in terms of spin degrees of freedom is proportional to 
  $\dd \langle E\rangle /\dd t\propto \langle J^x\rangle  \langle J^y\rangle .$
  In the steady state $\langle J^y\rangle =0$ and the system's energy is constant, indicating that all energy pumped from the external driving fields is completely lost through the dissipation of the cavity mode. However, on its way to stationarity, the spin component $\langle J^y\rangle $ oscillates around zero value, which means that with   period $2\omega_0$ energy is pumped in (negative $\langle J^y\rangle$) and out (positive $\langle J^y\rangle$) of the system, leading to the relaxation time $\tau=(\omega_y^{2}+\kappa^{2})/(\omega_{0}^2\kappa)\gg 1/\kappa$~\cite{PRADickeKeeling,PRLKeeling}.

In contrast, in the four-level model~\eqref{eq:model}, the
superradiance in one spin species  acts as an `effective drive' for the other, inducing an additional factor that slows down the relaxation. 
Here, the explicit breaking of the Hamiltonian symmetry results in the generation of non-stationary phases. 
It happens due to the competing conditions on the phases of the boson fields, $\phi_0,\ldots, \phi_3,$ set by $H$ [Eq.~\eqref{eq:hamiltonian_b}] and $H_s$ [Eq.~\eqref{eq:hamiltonian_s}].  
Relaxation in the four-level model~\eqref{eq:model} is conditioned from the temporal evolution of phases of the boson operators $\phi_{1,\ldots,3},$ whose interdependence slows down reaching a steady state, as it happens in constrained models~\cite{bhore2023deep,Riccardo2}. Such slow relaxation is crucial for the continuous read-out of the system's dynamics since spin precession can be easily captured at extensive timescales.

\subsection{Read-out}

We now relate  the dynamics of the auxiliary cavity  field $\langle a_y\rangle $ to the evolution of both the entanglement entropy $S_{\operatorname{vN}}$ and the condensate's microscopic degrees of freedom. An instance of such dynamics for parameters as in the red region in Fig.~\ref{fig:phase_diagram} ($\eta_s=0.3\eta_s^c$, $\eta=1.6\eta^c$) is shown in Fig.~\ref{fig:lattice_photon}. Here, the blue line in panel (a) shows the dynamics of the entanglement entropy, panel (b)  the dynamics of the populations of the two cavity modes, panel (c)  the dynamics of the phase of two cavity fields, and finally panel (d) shows snapshots of the condensate density at different times. Arrows in panel (d) indicate spin magnetization in the centers of the checkerboard lattice sites. 

The oscillations of the photon number $n_y=\langle a_y^{\dagger}a_y\rangle$ [panel(b)] and the phase $\phi_y=\arg(\langle a_y\rangle )$ [panel (c)] capture precession of the external pseudo-spin $\langle J^x\rangle, $ as $\langle a_y\rangle \approx -2\eta_s(\langle J_{01}^x\rangle +\langle J_{23}^x\rangle)/(\omega_y-i\kappa)$. When the phase of the cavity field changes by $\pm \pi,$ the real space density checkerboard lattice changes its parity [odd or even lattice sites are more occupied, see panel (d)]. Concomitantly, the system reaches the maximum value of the entanglement entropy [panel (a)]. At the same time,  the checkerboard lattice with period $\lambda=2\pi/k$ is formed not by modulation of the density but rather by the different orientations of the spin in the centers of even and odd sites.

When the phase of the field $\langle a_y\rangle $ gains $\pm\pi$ jump, the spatial density profile changes parity. 
At the same time, the increase of the photon number $n_y$ indicates a decrease of entanglement since the coupling $\eta_s$ tends to disentangle spin and momentum, while a decrease of $n_y$, on the other hand, indicates the developing of the spin-momentum correlations in the system. In this way, one can capture real-time oscillations of the entanglement entropy from the oscillations of the cavity field $\langle a_y\rangle.$

Finally, the fixed phase of the cavity field $a_z,$  $\phi_z=\arg(\langle a_z\rangle ),$ indicates the spontaneous symmetry breaking in $H$ [cf. Eq.~\eqref{eq:hamiltonian_b}]. In terms of  the atomic degrees of freedom, the fixed phase $\phi_z$ in panel (c) captures the absence of the mirror symmetry between  maximally entangled states, namely for two consecutive maximally entangled states, the spin lattices are exactly the same, without the symmetry under swapping  even and odd sites, cf. even panels in Fig.~\ref{fig:lattice_photon}(d). 


The heterodyne detection of two cavity modes $a_y$ and $a_z$ enables distinguishing different dynamical phases in the system in a non-destructive way. However, by itself, the coupling to the auxiliary cavity mode can change the steady state properties and, more importantly in the context of this paper, change the entanglement of the system compared to the single-mode model. In this regard, it is important to separate a range of couplings for which utilizing additional polarization preserves most of the entanglement and, at the same time, is sufficient to perform measurements. We dedicate the next Section to this aim. 

\section{Entanglement in the two photon fields model\label{sec:entanglement} }

In this Section, we revisit the system's various dynamical responses in terms of spin-momentum entanglement generation when both cavity modes contribute to the dynamics and identify parameter ranges suitable for non-invasive monitoring of the dynamics of collective observables. We show criteria to determine the range of parameters for which the auxiliary cavity mode creates a minimal backaction on the system's dynamics. For all extra details, we refer the reader to Appendix~\ref{sec:atom_only}.

\begin{figure}[t]
    \centering
\includegraphics[width=\linewidth]{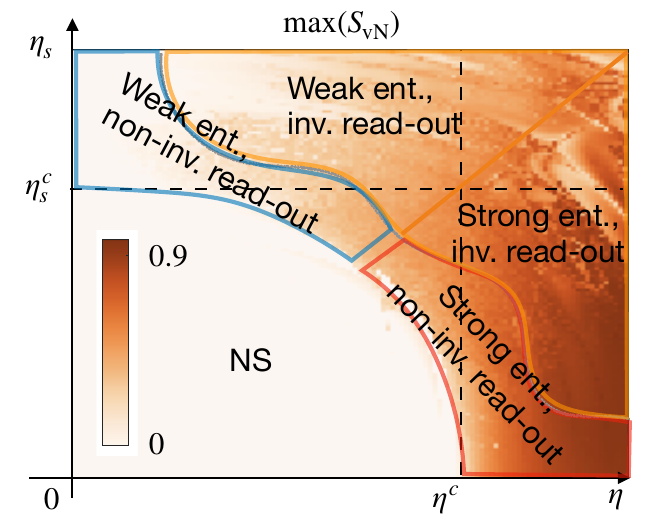}
    \caption{Maximum value of the entanglement entropy at late times as a function of the couplings $\eta $ and $\eta_s.$ 
    Parameters are the same as in Fig.~\ref{fig:phase_diagram}. The cartoon in Fig.~\ref{fig:entanglement0}(c) is sketched from the data of this figure.  \ox{Blue, yellow, and orange lines are added to distinguish different regimes in terms of the amount of spin-momentum entanglement and the invasiveness of the read-out procedure.} }
    \label{fig:fig3}
\end{figure}

The entanglement properties of the system are conditioned from the competition between $\eta$ [which couples spin-momentum pseudo-spins with cavity mode $a_z$ and tries to entangle spin and momentum] and $\eta_s$ [which couples momentum pseudo-spins $J$ with the cavity field $a_y$ and tends to maintain spin and momentum separable]. 
Fig.~\ref{fig:fig3} shows numerical data on the maximal entanglement $\max(S_{\operatorname{vN}})$  as a function of these `spin-momentum'  $\eta$ and `momentum' $\eta_s$ couplings.
Here, we maintain the same parameters as those in the phase diagram of Fig.~\ref{fig:phase_diagram}(a). The plot shows that  increasing $\eta $ induces stronger correlations between spin and momentum, while $\eta_s$ acts as a disentangling agent.

The interplay between $\eta$ and $\eta_s$ can  significantly alter not only the $\max(S_{\operatorname{vN}})$ but also the steady-state properties of the system along with the evolution of the $S_{\operatorname{vN}}$  compared to the $\eta_s=0$ case.  Fig.~\ref{fig:lattice_photon}(a) shows the evolution of the $S_{\operatorname{vN}}$ for weak, strong, and zero coupling $\eta_s$ to the auxiliary mode $a_y$. Compared to the unprobed model ($\eta_s=0$, black line), for weak values
 the photon-matter coupling slightly modifies the steady state and entanglement dynamics ($\eta_s=0.3\eta_s^c\ll\eta_s^c$, blue line), while for strong values,  it can alter dynamics of the $S_{\operatorname{vN}}$ significantly ($\eta_s=0.8\eta_s^c$, yellow line). 
 These regimes can be distinguished from the dynamics of the auxiliary field. For   weak couplings, the read-out is non-invasive, and the auxiliary field oscillates with the zero time average for timescales that significantly exceed the operational timescale of the experiment ($t_{\operatorname{op}}\propto 0.01$ s). In the experiment, these oscillations correspond to periodic changes in the phase of the auxiliary field. 
In the strong coupling regime, the read-out procedure is invasive, and  the phase of the auxiliary field becomes fixed after some time, $T,$ which is comparable with the operational timescale of the experiment. At this time, the explicitly broken symmetry of the Hamiltonian is restored, and the system starts evolving toward the SR state for both cavity fields. 

The restoration of the symmetry indicates the change in the steady state and, thus also entanglement properties of the system. Because of this, for non-destructive probing of the dynamics, it would be convenient  to work in a parameter regime  where  $T/t_{\operatorname{op}}\to \infty.$ Our numerics suggests that $T\propto\exp(\eta_s-\eta_s^c),$ revealing that symmetry restoration occurs more rapidly as the coupling to the auxiliary mode approaches the critical threshold, cf. Appendix~\ref{sec:atom_only}. As such, for non-destructive monitoring dynamics, it is essential to maintain the coupling to the auxiliary field significantly below the critical value. 
 
Combining the information on relaxation timescales 
with the amount of entanglement generated in  Fig.~\ref{fig:fig3} (see Appendix~\ref{sec:Redfield} for more details), it appears    that   in the  limit  of $\eta \gg\eta ^c $ and  $\eta_{s}\ll\eta_{s}^c $ \ox{(parameters region, separated by the red lines in Fig.~\ref{fig:fig3})}  highly entangled states are prepared, while the system keeps oscillating for long times, facilitating the reconstruction of oscillations of the entanglement entropy by measuring the auxiliary cavity field $\langle a_y\rangle$ (cf. Fig.~\ref{fig:lattice_photon}).
On the other hand, in the part of the phase diagram dominated by the coupling $\eta_s,$ there is no strong entanglement, albeit $\eta $   can still induce short-lived spin-momentum correlations \ox{(see parameters region, separated by the blue line in Fig.~\ref{fig:fig3}). Finally, when both couplings are high enough (see parameters region separated by yellow lines), probing the system's dynamics with the auxiliary cavity field alters dynamics significantly, and  $T$ is finite.}

By adjusting experimentally accessible parameters, such as couplings $\eta,~\eta_s$ and detuning $\omega_s$, one can  tune  the amplitude, time-average, and oscillation frequency of the entanglement entropy, thereby dynamically controlling the correlation between spin and momentum. In principle, the simplest approach  is to set $\omega_y/\kappa\to 0$, which ensures that pseudo-spins do not receive any feedback from $a_y$, and thus, all entanglement properties are solely determined by light-matter interactions contained in $H $.  In this case, the dissipation induces a phase shift of the  auxiliary cavity field, $\phi_\kappa^{y}=\tan ^{-1}\left(-\kappa / \omega_{y}\right)=\pi/2$~\cite{dogra2019dissipation}, making it imaginary, $\langle a_y\rangle+\langle a_y^{\dagger}\rangle=0,$ and reducing backaction of the field $a_y$ on the $H$, see Appendix~\ref{sec:atom_only}.
However, the effective model with $H_{\operatorname{tot}}$ [Eq.~\eqref{eq:model}] breaks down for these extreme conditions because the many-body description of the model, in this case, requires taking into account higher momentum modes. 

A more practical scenario is  when  the frequency of the auxiliary cavity photon $\omega_y$ is much higher than $\kappa.$ In this case, it is easier to maintain $\eta_s/\eta_s^c\ll 1,$ but have   cavity occupations $n_y $ large enough to continuously measure the collective momentum.

Furthermore, sizable stationary entanglement   can be preserved when 
$\omega_s= 0$.
 Here, the Hamiltonian gains additional symmetry under the exchange of ground and excited momentum levels of two spin sub-levels, $(\ket{0},\ket{1})\leftrightarrow(\ket{2},\ket{3})$. In this case, the induced as a result of the explicit symmetry breaking phase
 $\pm(\phi_{1}-\phi_3)$ does not evolve in time, and the explicitly broken symmetry can not be restored, see Appendix~\ref{sec:atom_only} for comprehensive discussion. 
For $\omega_s=0$ the maximal and time-averaged amount of entanglement remains similar to the one generated with one main cavity mode [see Fig.~\ref{fig:entanglemet_phase_d_1}(a)], besides a small dressing induced by $\eta_s$.

\section{CONCLUSIONS AND OUTLOOK} 
\label{sec:concl}
  
In this work, we have presented an experimentally feasible cavity QED platform featuring an effective four-level atomic description and shown that it manifests two intertwined self-organization transitions. This system serves as a minimal model wherein spontaneous symmetry breaking occurs in an all-to-all interacting spin model, concomitant with the formation of tunable spin-momentum entanglement. The controlled leakage of intra-cavity photons plays an important stabilization  role, as the resulting dissipative dynamics facilitate convergence towards the target entangled state in a manner resilient to imperfections in the system's couplings or initial state preparation. Extending the coupling scheme with an auxiliary cavity mode gives rise to persistent oscillations (due to explicit symmetry breaking) and facilitates real-time monitoring of the system dynamics, in particular, as a proxy for entanglement.

\bigskip

The tunable parameters of our model facilitate a straightforward extension to spin-exchange interactions, akin to the Tavis-Cummings model~\cite{soriente2018dissipation,Ferri,kirton2018superradiant}.
An interesting avenue for exploration lies in understanding how quantum correlations between spin and momentum can be continuously tuned as one transitions between the Tavis-Cummings and Dicke limits considered here.

We should note that a relation between multi-level atoms and entanglement has been previously reported both in cavity QED systems~\cite{Orioli_Dark,orioli2020subradiance,PhysRevResearch.5.013056,sundar2023squeezing}
 and photonic waveguides~\cite{pnas.1911467116,DChang2017designing,masson2022universality,sierra2022dicke,PhysRevLett.131.073602}. 
In these cases, entangled states can be hosted within the sub-radiant subspaces of the multi-level atoms, with level degeneracies being crucial for the build-up of quantum correlations. The mechanism is markedly different from ours, although considering a combination of the two setups could naturally lead to further interesting developments. 

Taking a broader perspective, one could investigate how different dynamical phases of matter routinely engineered in cavity QED would morph, when both spin and momentum degrees of freedom are optically addressed. Our analysis has focused on the superradiant phase transition as a paradigmatic case, but it would be intriguing to see whether entanglement properties of spin-momentum hybridized states can be manipulated as a response to periodic drives~\cite{Kessler1,Kessler2,reilly2023speeding} or in the context of dissipative-induced phase transitions~\cite{dogra2019dissipation,Phatthamon_continuousTC,Kessler_open_Dicke_nodel,buvca2019dissipation,soriente2018dissipation}.


\begin{acknowledgments}
The authors thank  Tilman Esslinger for fruitful discussions. O. C. is grateful to M. Stefanini and R.J. Valencia-Tortora for many useful discussions.
 This project has been supported by the Deutsche Forschungsgemeinschaft (DFG, German Research Foundation) – Project-ID 429529648 – TRR 306 QuCoLiMa ``Quantum Cooperativity of Light and Matter'')
 and by the QuantERA II Programme that has received funding from the European Union's Horizon 2020 Research and Innovation Programme under Grant Agreement No 101017733 (`QuSiED’) and by the DFG (project number 499037529). 
 We acknowledge funding from the Swiss National Science Foundation (project No. 212168 and IZBRZ2\_186312), from EU Horizon 2020 (European Research Council advanced grant TransQ, project no. 742579) and from the Swiss Secretariat for Education, Research and Innovation (SERI).
J.M. and O.C. acknowledge support from the Dynamics and Topology Centre funded by the State of Rhineland Palatinate. K.S. acknowledges funding from NSF EAGER-QAC-QCH award No. 2037687.  The authors gratefully acknowledge the computing time granted on the supercomputer MOGON 2 at Johannes Gutenberg-University Mainz (hpc.uni-mainz.de).
\end{acknowledgments}
\newpage
\appendix
\begin{widetext}

\section{\label{sec:derivation} Derivation of the Hamiltonian}

In this Appendix, we present a derivation of Hamiltonian $H_{\operatorname{tot}}$ for a setup depicted in Figs.~\ref{fig:1} and~\ref{fig:level_schemes_Hs}. The derivation of the Hamiltonian $H$ or $H_s$ can be obtained by setting $E_s=0$ or $E_b=E_r=0$ below respectively.

We start with the general Hamiltonian for the light-matter interaction problems, which reads
\begin{equation}
 H_{\operatorname{tot}}=H_{c}+H_a+H_{\text{int}},   
\end{equation}
where $H_c$ governs dynamics of the cavity mode, $H_a$ is single-atom Hamiltonian and $H_{\text{int}}$ describes interaction between atom and the cavity.  We specify the explicit form of $H_c,$ $H_a$ and $H_{\text{int}}$ in the considered setup and show under which assumptions the low-energy physics of the model can be simulated by Eq.~\eqref{eq:model}.

We immerse $^{87}$Rb atoms inside an optical cavity oriented along the $\mathbf{e}_{x}$ axis. The cavity has a single relevant frequency $\omega_{c}=2\pi\cdot 382.04685\text{ THz}$ with a decay rate of $\kappa\sim2\pi\cdot1.25\text{ MHz}$, and two polarizations $\mathbf{e}_y$ and $\mathbf{e}_z$ \ox{in the transverse plane.} We represent two corresponding  cavity  polarization modes by operators $ {a}_{ {y}}$ and $ {a}_{ {z}},$ respectively. The cavity Hamiltonian reads
\begin{equation}
    H_{c}=\hbar\omega_ya_y^{\dagger}a_y+\hbar\omega_z a_z^{\dagger}a_z.
\end{equation}

\ox{We apply a classical pump field with the standing-wave profile along $\mathbf{e}_z$ (perpendicular to the cavity axis) and 
 a Gaussian profile in the transverse directions, along $\mathbf{e}_{x}$ and $\mathbf{e}_{y}$.}
%
We consider a dispersive regime, in which pumping frequency is chosen out-of-resonance with the electron
transition $5^{2}S_{1/2}\to  5^{2}P_{1/2},~5^{2}P_{3/2}$. In this case, excited atomic states can be eliminated, and the resulting atomic Hamiltonian within the $F=1$ hyperfine manifold of the $5^{2}s_{1/2}$ level reads
\begin{equation}
    H_{a}=\frac{\mathbf{p}^2}{2M}+V_{\text{ext}}+\sum_{F,m_{F}}\hbar\omega_{F,m_{F}}|F,m_F\rangle\langle F,m_F|,
\end{equation}
where $\mathbf{p}$ is the momentum of the atom, $M$ is the atomic mass, $V_{\text{ext}}$ describes an external trapping potential, and the energy of atomic level $|F,m_F\rangle$ is $\hbar\omega_{F,m_{F}}.$ The sum in the last terms runs over all atomic levels in $F=1$ manifold. We apply a strong magnetic field $\mathbf{B}=-B\mathbf{e}_z$ along $z$ direction, which induces  first and second-order Zeeman splitting between levels with different magnetic numbers $m_F.$ Introducing internal spin operator $\mathbf{F}=\left(F^x,F^y,F^z\right)$, the atomic part of the Hamiltonian can be rewritten
\begin{equation}
    H_{a}=\frac{\mathbf{p}^2}{2M}+V_{\text{ext}}+\hbar\omega_z^{(1)}F^z+\hbar\omega_z^{(2)}(F^{z})^{2},
\end{equation}
where $\omega_z^{(1)}<0$ and $\omega_z^{(2)}>0$ are the first and the second order Zeeman splittings. 

The atoms are neutral and are much smaller than the wavelength of the optical light fields so that the light-matter interaction can be described in the dipole approximation
\begin{equation}
    {H}_{\text{int}}=-{\mathbf{d}}_{\text{lab}}{\mathbf{E}}_{\text{lab}},
\end{equation}
where the atomic dipole operator can be expanded in terms of the atom's internal states
\begin{equation}   {\mathbf{d}}_{\text{lab}}=\sum_{e,g}\mathbf{d}_{eg}\ket{e}\bra{g}+\text{h.c.},\quad  \mathbf{d}_{eg}=\bra{e}{\mathbf{d}}\ket{g}. 
\end{equation}

The total optical field $\mathbf{E}_{\text{lab}}$ is the sum of the classical pump field, ${\mathbf{E}}_{p}\left({\mathbf{r}}\right)$, and the cavity field, ${\mathbf{E}}_{c}\left({\mathbf{r}}\right)$. The classical pump field with the polarization along $\mathbf{e}_y$, has a standing-wave profile in the longitudinal direction ($\mathbf{e}_{z}$) and a
Gaussian profile in the  transverse direction (along $\mathbf{e}_{x}$, $\mathbf{e}_{y}$): 
\begin{equation}  \label{eq:classical_drive}\mathbf{E}_{p}\left({\mathbf{r}}\right)=\mathbf{e}_{y}\sum_{\beta=b,r,s}\frac{E_{\beta}}{2}f_{\beta}\left({\mathbf{r}}\right)e^{-i\omega_{\beta}t}+\text{h.c.},
\end{equation}
with  the mode function for each sideband is $f_{\beta}\left( {\mathbf{r}}\right)\equiv f\left( {\mathbf{r}}\right)=\exp\left({-\left({2{x}^{2}}/{w_{x}^{2}}-{2{y}^{2}}/{w_{y}^{2}}\right)}\right)\cos\left(k_{\beta}{z}\right)$. The widths of the transverse Gaussian profile are approximately $\omega_{x},\omega_{y}\approx25\text{ }\mu\text{m}$. The wave-vectors are $k_{\beta}={\omega_{\beta}}/{c}\approx{\omega_{p}}/{c}$ with $c$ denoting speed of light. 
 To induce the atomic transitions, discussed in Sec.~\ref{sec:Model} and Sec.~\ref{sec:monitoring_1}, we consider three laser drivers $\beta=b,r,s,$  
 with the sideband frequencies $\omega_{b(r)}$ such that detunings $\delta\omega_{\beta}=\omega_{\beta}-\omega_{p}$ are chosen to correspond to the differences in first-order Zeeman shifts (in the $F=1$ ground state manifold). In contrast, the third detuning is set to be zero, $\omega_s=(\omega_b+\omega_r)/2=\bar{\omega}$. In this case, different driving schemes operate with the same momentum states. 
 We limit our consideration to the case when $E_b=E_r\ne E_s.$

 The cavity field TEM00 mode of a Fabry-Perot cavity has a standing-wave profile in the transverse direction ($\mathbf{e}_x$) and a Gaussian profile in the other two directions along $\mathbf{e}_y $ and $\mathbf{e}_z$:
 \begin{equation}    \mathbf{E}_{c}\left( {\mathbf{r}}\right)=\mathbf{e}_{ {y}}E_{c}g_{ {y}}\left( {\mathbf{r}}\right) {a}_{ {y}}+\mathbf{e}_{ {z}}E_{c}g_{ {z}}\left( {\mathbf{r}}\right) {a}_{ {z}}+\text{h.c.}
 \end{equation}
 where the mode functions are $g_{{y},{z}}\left({\mathbf{r}}\right)\equiv g\left({\mathbf{r}}\right)=\exp\left(-({2\left({y}^{2}+ {z}^{2}\right))/{w_{c}^{2}}}\right)\cos\left(k_{c} {x}\right)$ with the Gaussian profile having a width of approximately $w_{c}\approx25\text{ }\mu\text{m}$. The wave-vectors are $k_{ {y}, {z}}\approx k_c=\omega_{c}/{c}$.

 As we are working in the dispersive regime, the excited atomic states can be eliminated using the Schrieffer-Wolff~\cite{haq2019explicit} transformation $ {H}\rightarrow e^{ {S}} {H}e^{- {S}}$, $\left[ {S}, {H}_{0}\right]=- {H}_{\text{int}}$, which results in the low-energy Hamiltonian
 \begin{equation}
\begin{aligned}
&  {H}_0=\frac{ {\mathbf{p}}^2}{2 M}+V_{\text{ext}}+\hbar \omega^{(1)}  {F}^z+\hbar \omega^{(2)}  ({F^z})^2+\hbar \omega_c^{ {y}} a_{ {y}}^{\dagger} a_{ {y}}+\hbar \omega_c^{ {z}} a_{ {z}}^{\dagger} a_{ {z}} \\
&  {H}_{\mathrm{int}}^s=\alpha_s f^2( {\mathbf{r}})\left|\sum_\beta \frac{E_\beta}{2} e^{i \delta \omega_\beta t}\right|^2+\alpha_s\left|E_c\right|^2 g^2( {\mathbf{r}})\left( {a}_{ {y}}^{\dagger}  {a}_{ {y}}+ {a}_{ {z}}^{\dagger}  {a}_{ {z}}\right) \\
& +\alpha_s\overbrace{\left(\mathbf{e}_y \cdot \mathbf{e}_{ {y}}\right)}^{=1} f( {\mathbf{r}}) g( {\mathbf{r}}) \sum_\beta \frac{E_\beta}{2}\left(e^{i \omega_\beta t}  {a}_{ {y}}+e^{-i \omega_\beta t}  {a}_{ {y}}^{\dagger}\right)+\alpha_s \overbrace{\left(\mathbf{e}_y \cdot \mathbf{e}_{ {z}}\right)}^{=0} f( {\mathbf{r}}) g( {\mathbf{r}}) \sum_\beta \frac{E_\beta}{2}\left(e^{i \omega_\beta t}  {a}_{ {z}}+e^{-i \omega_\beta t}  {a}_{ {z}}^{\dagger}\right) \\
&  {H}_{\mathrm{int}}^v=-i \frac{\alpha_v}{2 F} g( {\mathbf{r}}) f( {\mathbf{r}}) \sum_\beta \frac{E_c E_\beta}{2}\left[\left( {a}_{ {y}} e^{i \omega_\beta t}- {a}_{ {y}}^{\dagger} e^{-i \omega_\beta t}\right)\overbrace{\left(\mathbf{e}_y \times \mathbf{e}_{ {y}}\right)}^{=0} \cdot  {\mathbf{F}}+\left( {a}_{ {z}} e^{i \omega_\beta t}- {a}_{ {z}}^{\dagger} e^{-i \omega_\beta t}\right)\overbrace{\left(\mathbf{e}_y \times \mathbf{e}_{ {z}}\right)}^{=\mathbf{e}_x} \cdot  {\mathbf{F}}\right] \\
& -\underbrace{i \frac{\alpha_v}{2 F} E_c^2 g^2( {\mathbf{r}})\left( {a}_{ {y}}^{\dagger}  {a}_{ {z}}- {a}_{ {z}}^{\dagger}  {a}_{ {y}}\right)\left(\mathbf{e}_{ {y}} \times \mathbf{e}_{ {z}}\right) \cdot \overbrace{ {\mathbf{F}}}^{\propto e^{ \pm i \omega t}} }_{=0, \text { because of resonance conditions }} 
\end{aligned}
\end{equation}
For the transitions in multi-level atoms, it is convenient to account for selection rules using polarizabilities. \ox{In the above equation, $\alpha_{s,v}$ are scalar and vector polarizabilities of the atoms, which are components of the rank-2 tensor $\alpha_{i,j}=\sum_{g,g^{\prime}}\sum_{e}\langle g|d_i|e\rangle \langle e|d_j^{\dagger}|g^{\prime}\rangle |g\rangle\langle g^{\prime}|/(\hbar \Delta_e)=\alpha_{\mathrm{s}}  {\mathrm{I}} \delta_{i j}-i {\alpha_{\mathrm{v}}}/({2 \mathrm{~F}}) \epsilon_{i j k}+\ldots $. Here, diagonal components, proportional to $\alpha_s$, describe the process when the spin of the atom remains unchanged, while vectorial non-diagonal components, proportional to $\alpha_v$, describe the transition of the spin state of the atom after the two-photon process.}
The sum runs over all allowed transitions (see Ref.~\cite{steck2001rubidium}), $d_i=(\mathbf{d}\mathbf{e}_i )$ is the $i-$th component of the atomic dipole moment, $i=\{x,y,z\}$, and the detuning of the driving field from the resonance frequency is $\Delta_e\approx \omega_p-\omega_e.$

Finally, we move to a frame rotating with the classical pump  frequency: $ {H}_{\text{lab}}\rightarrow {H}= {U}_{p}^{\dagger} {H}_{\text{lab}} {U}_{p}- {H}_{p}$ where $ {U}_{p}=\exp(-{i {H}_{p}t}/{\hbar})$ and $ {H}_{p}=\sum_{e}\hbar\omega_{p}\ket{e}\bra{e}+\hbar\omega_{p}\left( {a}_{ {y}}^{\dagger} {a}_{ {y}}+ {a}_{ {z}}^{\dagger} {a}_{ {z}}\right)$. The rotating-wave approximation brings us to the time-independent single-body Hamiltonian
\begin{equation}
    H_{\operatorname{tot}}=H_{a}+H_c+H_s+H_v,
\end{equation}
\begin{equation}
    H_a=\frac{\mathbf{p}^2}{2 M}+V_{\text{ext}}+\hbar \delta^{(1)} F^z+\hbar \omega^{(2)} (F^{z})^{2},
\end{equation}
\begin{equation}
    H_c=-\Delta_y a_y^{\dagger}a_y-\Delta_z a_z^{\dagger}a_z,
\end{equation}
\begin{equation}
   H_s=\frac{\alpha_s}{4} f^2(\mathbf{r})\left(E_s^2+E_b^2+E_r^2\right)+\alpha_s E_0^2g^2(\mathbf{r})\left(a_y^{\dagger}a_y+a_z^{\dagger}a_z\right)+\frac{\alpha_s}{2} f(\mathbf{r}) g(\mathbf{r}) E_0 E_s(a_y+a_y^{\dagger})
\end{equation}
\begin{equation}
    H_v=\frac{\alpha_v}{4}E_0 E_{b} f(\mathbf{r})g(\mathbf{r})(a_z+a_z^{\dagger })F^x
\end{equation}
\noindent    where
$\delta^{(1)}=\omega_z^{(1)}+(\omega_b-\omega_r)/2,$ $\Delta_{y,z}=\bar{\omega}-\omega_{y,z}<0.$ We have also applied a transformation $a_z\to i a_z$ to get rid of the minus sign in $H_v.$ 
The first term in $H_s$ describes the attractive potential created by the transverse driving fields, the second term describes the dispersive shift to the cavity detuning, and the last term produces the Bragg transition within the same atomic level. The vectorial interaction describes the Raman process when the transition happens between nearing sub-levels of the ground-state manifold. 

Let us consider a case when the magnetic field is strong. Specifically, let the second order Zeeman shift $\omega^{(2)}\propto 1~\text{MHz},$ and thus the resonant conditions for transition $m_F=1\leftrightarrow m_F=0 $ are out-of-resonance for transition $m_F=0\leftrightarrow m_F=-1.$ In this case, if we prepare the initial state as a mixture of particles at levels with $m_F=1,0,$ the dynamics will be restricted to these two atomic levels for the typical operational times of the experiment.
Thus, we can limit our consideration to the dynamics between two neighboring spin levels, defining the many-body spinor field operator
\begin{equation}
    \Psi(\mathbf{r})=(0,\psi_{\uparrow},\psi_{\downarrow})^T, 
\end{equation}
which satisfies standard bosonic commutation relations $[ \psi_{\sigma}(\mathbf{r}), \psi_{\sigma^{\prime}}^{\dagger }(\mathbf{r}^{\prime})]=\delta_{\sigma,\sigma^{\prime}}\delta(\mathbf{r}-\mathbf{r}^{\prime}),$ $[ \psi_{\sigma}(\mathbf{r}), \psi_{\sigma^{\prime}}(\mathbf{r}^{\prime})]=0,$ where $\sigma,\sigma^{\prime}=\uparrow,\downarrow.$
The $N-$body Hamiltonian reads
\begin{equation}\label{eq:BEC}
\begin{aligned}
  H_{\operatorname{tot}}&=H_{c}+\int \dd \mathbf{r}\Psi^{\dagger}(\mathbf{r})\left(H_a+H_s+H_v\right)\Psi(\mathbf{r})=\\
  &=H_{c}+\int \dd \mathbf{r}\left[\frac{\alpha_s}{4} \left(E_s^2+E_b^2+E_r^2\right)f^2(\mathbf{r})+\alpha_s E_0^2 \left(a_y^{\dagger} a_y+a_z^{\dagger} a_z\right)g^2(\mathbf{r})\right]\left(\psi^{\dagger}_{\uparrow}(\mathbf{r})\psi_{\uparrow}(\mathbf{r})+\psi^{\dagger}_{\downarrow}(\mathbf{r})\psi_{\downarrow}(\mathbf{r})\right)\\
  &+\int \dd \mathbf{r}\psi^{\dagger}_{\uparrow}(\mathbf{r})\left(-\frac{\hbar^2\nabla^2}{2 M}+V_{\mathrm{ext}}\right)\psi_{\uparrow}(\mathbf{r})+\int \dd \mathbf{r}\psi^{\dagger}_{\downarrow}(\mathbf{r})\left(-\frac{\hbar^2\nabla^2}{2 M}+V_{\mathrm{ext}}\right)\psi_{\downarrow}(\mathbf{r})+\\
  &+\frac{\hbar \delta^{(1)}}{2}\int \dd \mathbf{r} \left(\psi^{\dagger}_{\uparrow}(\mathbf{r})\psi_{\uparrow}(\mathbf{r})-\psi^{\dagger}_{\downarrow}(\mathbf{r})\psi_{\downarrow}(\mathbf{r})\right)+\frac{\hbar \omega^{(2)}}{4} \int \dd \mathbf{r}\left(\psi^{\dagger}_{\uparrow}(\mathbf{r})\psi_{\uparrow}(\mathbf{r})+\psi^{\dagger}_{\downarrow}(\mathbf{r})\psi_{\downarrow}(\mathbf{r})\right)\\
  &+\frac{\alpha_s}{2}E_0 E_s\left(a_y+a_y^{\dagger}\right)\int \dd \mathbf{r} f(\mathbf{r}) g(\mathbf{r})\left(\psi^{\dagger}_{\uparrow}(\mathbf{r})\psi_{\uparrow}(\mathbf{r})+\psi^{\dagger}_{\downarrow}(\mathbf{r})\psi_{\downarrow}(\mathbf{r})\right) +\\
  &+
  \frac{\alpha_v}{8} E_0 E_{b}\left(a_z+a_z^{\dagger}\right)\int \dd \mathbf{r}f(\mathbf{r}) g(\mathbf{r})\left(\psi^{\dagger}_{\uparrow}(\mathbf{r}) \psi_{\downarrow}(\mathbf{r})+\psi^{\dagger}_{\downarrow}(\mathbf{r})  \psi_{\uparrow}(\mathbf{r})\right).
\end{aligned}
\end{equation}
To derive an extended Dicke Hamiltonian, we should further restrict the Hilbert space of the model by considering only the two lowest momentum states of the model for both spinor components. In this approximation the many-body wave function reads $\Psi(\mathbf{r})=(0,\phi_0(\mathbf{r}) c_0+\phi_1(\mathbf{r})c_1,\phi_0 c_2(\mathbf{r})+\phi_1(\mathbf{r})c_3)^T.$ Here, $c_i$ are the annihilation operators of the corresponding atomic modes, $[c_i,c_j^{\dagger}]=\delta_{i,j}, ~[c_i,c_j]=0$, and $\phi_1(\mathbf{r})=\mathcal{N}\cos kx\cos kz \phi_0(\mathbf{r}),$ where $\mathcal{N}$ accounts for the correct normalization. In this notation, operators $c_0$ and $c_2$ correspond to the ground momentum states, while operators $c_1$ and $c_2$ correspond to the excited momentum states. After integrating over all space, the many-body Hamiltonian reads
\begin{equation}
  \begin{aligned}
H_{\operatorname{tot}}= & -\hbar\tilde{\Delta}_y a_y^{\dagger} a_y+\hbar \omega_{0}\left( {c}_1^{\dagger}  {c}_1+ {c}_3^{\dagger}  {c}_3\right)+\hbar\eta_s\left( {a}_{ {y}}+ {a}_{ {y}}^{\dagger}\right)\left( {c}_0^{\dagger}  {c}_1+ {c}_1^{\dagger}  {c}_0+ {c}_2^{\dagger}  {c}_3+ {c}_3^{\dagger}  {c}_2\right) \\
& -\hbar\tilde{\Delta}_z a_z^{\dagger} a_z+\hbar\omega_s\left( {c}_2^{\dagger}  {c}_2+ {c}_3^{\dagger}  {c}_3\right)+\hbar\eta \left( {a}_{ {z}}+ {a}_{ {z}}^{\dagger}\right)\left( {c}_1^{\dagger}  {c}_2+ {c}_0^{\dagger}  {c}_3+ {c}_2^{\dagger}  {c}_1+ {c}_3^{\dagger}  {c}_0\right),
\end{aligned}  
\end{equation}
where the cavity detuning is dressed via  the dynamic (dispersive) shift $\tilde{\Delta}_{y,z}=\Delta_{y,z}-\alpha_s E_0^2  N \mathcal{I}/\hbar$, the level splitting between ground and excited momentum states is equal $\omega_0=2\omega_{\text{rec}}$, $\omega_s=\delta^{(1)}$, and coupling constants read as follows: $\hbar\eta_{s}={\alpha_s}E_0 E_{s}\mathcal{M}/2$, $\hbar \eta =-{\alpha_v} E_0 E_{b}\mathcal{M}/{4\sqrt{2}}$, where $\mathcal{I}=\int \dd\mathbf{r} g^2(\mathbf{r})\phi_0^2(\mathbf{r}),$ $\mathcal{M}=\int \dd\mathbf{r}f(\mathbf{r})f(\mathbf{r})\psi_1(\mathbf{r})\psi_0(\mathbf{r})$. We have omitted standing-wave potential  ${\alpha_s}\left(E_s^2+E_b^2+E_r^2\right)\int \dd \mathbf{r}f^2(\mathbf{r})\phi_0^2(\mathbf{r})/4$ above as it only contributes to the higher momentum state and does not qualitatively modify the appearance of the phase transition discussed in the main text in Fig.~\ref{fig:phase_diagram}. 

Finally, one can introduce pseudo-spin operators according to Eq.~\eqref{eq:sigma}, which brings the Hamiltonian to the following form:
\begin{equation}\label{eq:Htot}
    \begin{aligned}
H_{\operatorname{tot}} &= \omega_y a_y^{\dagger} a_y+ \omega_{0}\left(J_{01}^z+J_{23}^z\right)+\eta_s\left(a_y+a_y^{\dagger}\right)\left(J_{01}^{+}+J_{23}^{+}+J_{01}^{-}+J_{23}^{-}\right)+ \\
& +\omega_z a_z^{\dagger} a_z+\omega_s\left(S_{12}^z+S_{03}^z\right)+\eta \left(a_z+a_z^{\dagger}\right)\left(S_{12}^{+}+S_{03}^{+}+S_{12}^{-}+S_{03}^{-}\right),
\end{aligned}
\end{equation}
where we set $\omega_{y,z}=-\tilde{\Delta}_{y,z}>0$ and have normalized the Hamiltonian by $\hbar\omega_{\text{rec}}$ so that $\omega_0=2$ above. \ox{One should take into account that due to the particle conservation, one has $J_{01}^z+J_{23}^z=S_{03}^z-S_{12}^z$. Additionally, when $\eta=0$, the term $\omega_s\left(S_{12}^z+S_{03}^z\right)$ can be omitted because the process $\eta_s\left(a_y+a_y^{\dagger}\right)\left(J_{01}^{+}+J_{23}^{+}+J_{01}^{-}+J_{23}^{-}\right)$ preserves the number of particles in the upper and lower spinor components, and thus the contribution to the energy of the system from the $\omega_s\left(S_{12}^z+S_{03}^z\right)$ remains constant during the whole course of the dynamics.}


\section{\label{sec:EoM}Equations of motion }
In this Appendix, we derive mean-field equations of motion from the Hamiltonian in~\eqref{eq:Htot}. Note that for the remainder of the Appendices, we omit $\langle\cdot\rangle$ for expectation values of observables for simplicity. 

The mean-field equations of motion can be easily derived from the Lindblad master equation~\eqref{eq:Lindblad} and read  
\begin{equation}\label{eq:eom}
    \begin{aligned}
        \dot{a}_{y}	&=-i(\omega_{y}-i\kappa)a_{ y}-i\eta_{s}\left( {c}_{0}^{\dagger} {c}_{1}+ {c}_{1}^{\dagger} {c}_{0}+ {c}_{2}^{\dagger} {c}_{3}+ {c}_{3}^{\dagger} {c}_{2}\right)\\
\dot{a}_{z}	&=-i(\omega_{z}-i\kappa)a_{ z}-i\eta \left( {c}_{1}^{\dagger} {c}_{2}+ {c}_{0}^{\dagger} {c}_{3}+ {c}_{2}^{\dagger} {c}_{1}+ {c}_{3}^{\dagger} {c}_{0}\right)\\
\dot{c}_{0}	&=-i\eta \left( {a}_{ z}+ {a}_{ z}^{\dagger}\right) {c}_{3}-i\eta_{s}\left( {a}_{ y}+ {a}_{ y}^{\dagger}\right) {c}_{1}\\
\dot{c}_{1}	&=-i\omega_{0} {c}_{1}-i\eta \left( {a}_{ z}+ {a}_{ z}^{\dagger}\right) {c}_{2}-i\eta_{s}\left( {a}_{ y}+ {a}_{ y}^{\dagger}\right) {c}_{0}\\
\dot{c}_{2}	&=-i\omega_{s} {c}_{2}-i\eta \left( {a}_{ z}+ {a}_{ z}^{\dagger}\right) {c}_{1}-i\eta_{s}\left( {a}_{ y}+ {a}_{ y}^{\dagger}\right) {c}_{3}\\
\dot{c}_{3}	&=-i\left(\omega_{s}+\omega_{0}\right) {c}_{3}-i\eta \left( {a}_{ z}+ {a}_{ z}^{\dagger}\right) {c}_{0}-i\eta_{s}\left( {a}_{ y}+ {a}_{ y}^{\dagger}\right) {c}_{2}\\
    \end{aligned}
\end{equation}

In terms of pseudo-spin degrees of freedom~\eqref{eq:sigma}, the equations of motion take the following form
\begin{equation}\label{eq:eom_spins}
\begin{aligned}
\frac{d J_{01}^{-}}{d t} & =-2 i \omega_0 J_{01}^{-}+2 i \eta_s\left(a_y+a_y^{\dagger}\right) J_{01}^z+ i \eta \left(a_z+a_z^{\dagger}\right)\left(\mathcal{T}_{13}^{+}-\mathcal{T}_{02}^{-}\right) \\
\frac{d J_{23}^{-}}{d t} & =-2 i \omega_0 J_{23}^{-}+2 i \eta_s\left(a_y+a_y^{\dagger}\right) J_{23}^z+ i \eta \left(a_z+a_z^{\dagger}\right)\left(\mathcal{T}_{13}^{-}-\mathcal{T}_{02}^{+}\right) \\
\frac{d J_{01}^z}{d t} & =i \eta_s\left(a_y+a_y^{\dagger}\right)\left(J_{01}^{-}-J_{01}^{+}\right)+i \eta  / 2\left(a_z+a_z^{\dagger}\right)\left(S_{12}^{+}-S_{12}^{-}+S_{03}^{-}-S_{03}^{+}\right) \\
\frac{d J_{23}^z}{d t} & =i \eta_s\left(a_y+a_y^{\dagger}\right)\left(J_{23}^{-}-J_{23}^{+}\right)+ i \eta  / 2\left(a_z+a_z^{\dagger}\right)\left(S_{12}^{+}-S_{12}^{-}+S_{03}^{-}-S_{03}^{+}\right) \\
\frac{d S_{12}^{-}}{d t} & =2 i\left(\omega_0-\omega_s\right) S_{12}^{-}+2 i \eta \left(a_z+a_z^{\dagger}\right) S_{12}^z+i \eta_s\left(a_y+a_y^{\dagger}\right)\left(\mathcal{T}_{02}^{-}-\mathcal{T}_{13}^{-}\right) \\
\frac{d S_{03}^{-}}{d t} & =-2 i\left(\omega_0+\omega_s\right) S_{03}^{-}+2 i \eta \left(a_z+a_z^{\dagger}\right) S_{03}^z+i \eta_s\left(a_y+a_y^{\dagger}\right)\left(\mathcal{T}_{13}^{-}-\mathcal{T}_{02}^{-}\right) \\
\frac{d S_{12}^z}{d t} & =i \eta \left(a_z+a_z^{\dagger}\right)\left(S_{12}^{-}-S_{12}^{+}\right)+i \eta_s / 2\left(a_y+a_y^{\dagger}\right)\left(-J_{01}^{-}+J_{01}^{+}-J_{23}^{-}+J_{23}^{+}\right) \\
\frac{d S_{03}^z}{d t} & =i \eta \left(a_z+a_z^{\dagger}\right)\left(S_{03}^{-}-S_{03}^{+}\right)+ i \eta_s / 2\left(a_y+a_y^{\dagger}\right)\left(J_{01}^{-}-J_{01}^{+}+J_{23}^{-}-J_{23}^{+}\right) \\
\frac{d \mathcal{T}_{13}^{-}}{d t} & =-2 i \omega_s \mathcal{T}_{13}^{-}+i \eta \left(a_z+a_z^{\dagger}\right)\left(J_{23}^{-}-J_{01}^{+}\right)+i \eta_s\left(a_y+a_y^{\dagger}\right)\left(S_{03}^{-}-S_{12}^{-}\right) \\
\frac{d \mathcal{T}_{02}^{-}}{d t} & =-2 i \omega_s \mathcal{T}_{02}^{-}+ i \eta \left(a_z+a_z^{\dagger}\right)\left(J_{23}^{+}-J_{01}^{-}\right)+i \eta_s\left(a_y+a_y^{\dagger}\right)\left(S_{12}^{-}-S_{03}^{-}\right) \\
\frac{d \mathcal{T}_{13}^z}{d t} & =i \eta  / 2\left(a_z+a_z^{\dagger}\right)\left(S_{03}^{-}-S_{03}^{+}+S_{12}^{-}-S_{12}^{+}\right)+i \eta_s / 2\left(a_y+a_y^{\dagger}\right)\left(J_{23}^{-}-J_{23}^{+}-J_{01}^{-}+J_{01}^{+}\right) \\
\frac{d \mathcal{T}_{02}^z}{d t} & =i \eta  / 2\left(a_z+a_z^{\dagger}\right)\left(S_{03}^{-}-S_{03}^{+}+S_{12}^{-}-S_{12}^{+}\right)+i \eta_s / 2\left(a_y+a_y^{\dagger}\right)\left(J_{23}^{+}-J_{23}^{-}+J_{01}^{-}-J_{01}^{+}\right)
\end{aligned}
\end{equation}

Note that on the mean-field level, both~\eqref{eq:eom} and \eqref{eq:eom_spins} govern identical dynamics when the system is initially prepared in the coherent state. However, if the initial state contains higher-order correlations, one needs to consider higher-order corrections (i.e., cumulants expansion or similar methods) to capture dynamics accurately~\cite{Riccardo}. 

\bigskip

Finally, the equations of motion on the mean-field level can be derived without truncation over momentum states, starting from Hamiltonian~\eqref{eq:BEC}, $i\hbar\partial_t \Psi(\mathbf{r},t)=H_{\operatorname{tot}}(\mathbf{r})\Psi(\mathbf{r},t),$ which results into the following equations of motion 

\begin{equation}\label{eq:eomBEC}
    \begin{aligned}
        i\hbar \dfrac{d\psi_{\uparrow}(\mathbf{r})}{dt}&\approx 
        \frac{\hbar \delta^{(1)}}{2}\psi_{\uparrow}(\mathbf{r})
        -\frac{\hbar^2 \nabla^2}{2 M}\psi_{\uparrow}(\mathbf{r})+\overbrace{\frac{\alpha_s}{2} E_0 E_s}^{\tilde{\eta}_s}\left(a_y+a_y^{\dagger}\right) f(\mathbf{r}) g(\mathbf{r})\psi_{\uparrow}(\mathbf{r})+
        \overbrace{\frac{\alpha_v}{8} E_0 E_b}^{\tilde{\eta}}\left(a_z+a_z^{\dagger}\right) f(\mathbf{r}) g(\mathbf{r})\psi_{\downarrow}(\mathbf{r})
        \\
        i\hbar\dfrac{d\psi_{\downarrow}(\mathbf{r})}{dt}&\approx -\frac{\hbar \delta^{(1)}}{2}\psi_{\downarrow}(\mathbf{r})
        -\frac{\hbar^2 \nabla^2}{2 M}\psi_{\downarrow}(\mathbf{r})+{\frac{\alpha_s}{2} E_0 E_s}\left(a_y+a_y^{\dagger}\right) f(\mathbf{r}) g(\mathbf{r})\psi_{\downarrow}(\mathbf{r})+
        {\frac{\alpha_v}{8} E_0 E_b}\left(a_z+a_z^{\dagger}\right) f(\mathbf{r}) g(\mathbf{r})\psi_{\uparrow}(\mathbf{r})
        \\
         i\hbar\dfrac{da_y}{dt}&\approx \hbar (\omega_y-i\kappa) a_y+ \frac{\alpha_s}{2} E_0 E_s\int \mathrm{d} \mathbf{r} f(\mathbf{r}) g(\mathbf{r})\left(\psi_{\uparrow}^{\dagger}(\mathbf{r}) \psi_{\uparrow}(\mathbf{r})+\psi_{\downarrow}^{\dagger}(\mathbf{r}) \psi_{\downarrow}(\mathbf{r})\right)\\
         i\hbar\dfrac{da_z}{dt}&\approx \hbar (\omega_z-i\kappa) a_z+ 
         \frac{\alpha_v}{8} E_0 E_b \int \mathrm{d} \mathbf{r} f(\mathbf{r}) g(\mathbf{r})\left(\psi_{\uparrow}^{\dagger}(\mathbf{r}) \psi_{\downarrow}(\mathbf{r})+\psi_{\downarrow}^{\dagger}(\mathbf{r}) \psi_{\uparrow}(\mathbf{r})\right)
         \\
    \end{aligned}
\end{equation}

To retrieve dynamics at short times one can sufficiently simplify equations by eliminating cavity fields, substituting $a_y\approx-{\alpha_s}{} E_0 E_s \int \mathrm{d} \mathbf{r} f(\mathbf{r}) g(\mathbf{r})\left(\psi_{\uparrow}^{\dagger}(\mathbf{r}) \psi_{\uparrow}(\mathbf{r})+\psi_{\downarrow}^{\dagger}(\mathbf{r}) \psi_{\downarrow}(\mathbf{r})\right)/(2\hbar(\omega_y-i\kappa))$ and $a_z\approx -{\alpha_v}{} E_0 E_b \int \mathrm{d} \mathbf{r} f(\mathbf{r}) g(\mathbf{r})\left(\psi_{\uparrow}^{\dagger}(\mathbf{r}) \psi_{\downarrow}(\mathbf{r})+\psi_{\downarrow}^{\dagger}(\mathbf{r}) \psi_{\uparrow}(\mathbf{r})\right)/(8\hbar(\omega_z-i\kappa))$. Choosing periodic boundary conditions for field $\Psi(\mathbf{r})$, the resultant equations of motion can be further efficiently evaluated with the split-step Fourier transform method~\cite{sinkin2003optimization}. 
 
Both equations~\eqref{eq:eom} and ~\eqref{eq:eomBEC} describe the similar dynamical behavior of the system and transition from the normal to superradiant phase with the subsequent population of the momentum states $\ket{1}_m$ (note, that Eq.~\eqref{eq:eomBEC} also captures population of the momentum states $\ket{\pm 2 k,0,0}$ and $\ket{0,0,\pm 2 k}$, however for the most of the parameters the fraction of atoms there can be neglected).   
However, to study long-time dynamics, evaluation of ~\eqref{eq:eom} can be done much more efficiently, with much less computational cost. The further simplification via proper elimination of the cavity fields is described in the following Appendix.

 \section{Redfield equations\label{sec:Redfield}}

To evaluate dynamics at late times, we also adiabatically eliminate dissipative cavity modes $a_y$ and $a_z$ and study the atom-only model. This procedure is justified by the separation of scales between cavity detunings/decay rates and atomic frequencies, which differ by two to three orders of magnitude.  By applying the Schrieffer-Wolff transformation, the atom and photon modes can be decoupled, resulting in the effective atom-only description of the model. Following calculations in Refs.~\cite{Jager,atom_only}, we derive the following expressions for the effective fields $\alpha_y$ and $\alpha_z$
\begin{equation}\label{eq:redfield}
\begin{aligned}
 \alpha_{y}&=\dfrac{\eta_{s}\left(c_{1}^{\dagger}c_{0}+c_{3}^{\dagger}c_{2}\right)}{-\omega_{y}-\omega_{0}+i\kappa}+\dfrac{\eta_{s}\left(c_{0}^{\dagger}c_{1}+c_{2}^{\dagger}c_{3}\right)}{-\omega_{y}+\omega_{0}+i\kappa}\\
 \alpha_{z}&=\frac{\eta c_{3}^{\dagger}c_{0}}{-\omega_{z}-\left(\omega_{s}+\omega_{0}\right)+i\kappa}+\frac{\eta c_{0}^{\dagger}c_{3}}{-\omega_{z}+\left(\omega_{s}+\omega_{0}\right)+i\kappa}+\frac{\eta c_{2}^{\dagger}c_{1}}{-\omega_{z}+\left(\omega_{0}-\omega_{s}\right)+i\kappa}+\frac{\eta c_{1}^{\dagger}c_{2}}{-\omega_{z}-\left(\omega_{0}-\omega_{s}\right)+i\kappa}.
 \end{aligned}
\end{equation}
On the mean-field level, the effective equations of motion can be derived by substituting into Eqs.~\ref{eq:eom} $(\alpha_y,\alpha_z)$ instead of boson fields $(a_y, a_z).$ 
The atom-only model allows us to investigate long-time dynamics and numerically explore relaxation processes. However, it is important to note that this model operates correctly when both couplings are ramped up gradually. Abrupt changes in coupling can excite high-energy excitations in the model, which are not accounted for by the Redfield equation.

To evaluate the long-time dynamical response in Fig.~\ref{fig:phase_diagram}(a) (and also in Fig.~\ref{fig:fig3}), we initialize the system in the normal state and then ramp up both coupling during $\approx 0.002$ s, which is slow enough to make Redfield description of the dynamics valid and, at the same time, fast enough to excite non-stationary phases. We then compare dynamical properties of the system ($\langle a_y\rangle$, $\langle a_z\rangle$) after the ramp at $t\approx 0.01$ s and at late times $t\approx 0.4$ s to distinguish between phases with the explicitly broken symmetry (which suit for non-invasive dynamics monitoring) and phases with explicitly broken symmetry, which identify strong coupling regime and invasive probing of the dynamics. 


\section{\label{app:dicke} Analytical calculation of the steady state}
Both models~\eqref{eq:hamiltonian_s} and~\eqref{eq:hamiltonian_b} take the form of the Dicke model and thus undergo a phase transition associated with the spontaneous breaking of $\mathbb{Z}_2$ symmetry. The phases associated with this symmetry breaking are the normal phase, in which all  spins are polarized long $z$ direction and occupation of the cavity photon is zero, and the superradiant phase, in which spins develop a non-zero $x$ component, with the cavity occupation taking a non-zero value. The solution for each case can be derived as a stable stationary  state of Eqs.~\eqref{eq:eom_spins}. As an illustration, let us examine the case when one of the couplings is equal to zero. 

\subsection{$\eta =0$ case}
In this case, the critical coupling is equal to $\eta_{s}^{c}=\sqrt{{\omega_{0}\left(\omega_{y}^{2}+\kappa^{2}\right)}/({4\omega_{y}})}$ and the solution in the SR phase read $J_{01}^z=-\mu\eta_s^{c2}/(2\eta_s^2),$ $J_{23}^z=-(1-\mu)\eta_s^{c2}/(2\eta_s^2),$ $J^x=\pm \sqrt{1/4-J^{z2}}$, $a_y=-\eta_s/(\omega_y-i\kappa)\sqrt{1-\eta_s^{c4}/\eta_s^4}.$  
Interestingly, in this case, the spins $\mathcal{T}, ~S$ are not stationary but instead can precess according to the equations of motion
\begin{equation}
    \begin{aligned}
\frac{d S_{12}^{-}}{d t} & =2 i\left(\omega_0-\omega_s\right) S_{12}^{-}+i \eta_5\left(a_y+a_y^{\dagger}\right)\left(\mathcal{T}_{02}^{-}-\mathcal{T}_{13}^{-}\right) \\
\frac{d S_{03}^{-}}{d t} & =-2 i\left(\omega_0+\omega_s\right) S_{03}^{-}+1 i \eta_5\left(a_y+a_y^{\dagger}\right)\left(\mathcal{T}_{13}^{-}-\mathcal{T}_{02}^{-}\right) \\
\frac{d \mathcal{T}_{13}^{-}}{d t} & =-2 i \omega_s \mathcal{T}_{13}^{-}+1 i \eta_5\left(a_y+a_y^{\dagger}\right)\left(S_{03}^{-}-S_{12}^{-}\right) \\
\frac{d \mathcal{T}_{02}^{-}}{d t} & =-2 i \omega_s \mathcal{T}_{02}^{-}+1 i \eta_5\left(a_y+a_y^{\dagger}\right)\left(S_{12}^{-}-S_{03}^{-}\right)
\end{aligned}
\end{equation}
In the normal phase, the frequency of the precession is equal to $\pm 2\omega_s,$ while in the SR phase, additional dressing from the interaction with the cavity mode $a_y$ takes place
\begin{equation}
\Omega=-2\omega_{s}\pm \frac{4\sqrt{\omega_{0}^{2}\left(\omega_y^{2}+\kappa^{2}\right)^{2}/4-\omega_y^{2}\left(\eta_{s}^{c4}-\eta_s^{4}\right)}}{\left(\omega_y^{2}+\kappa^{2}\right)}.
\end{equation}
These dynamics stem from the fact that three species of the pseudo-spins in the model governed by $H_{\operatorname{tot}}$ in Eq.~\eqref{eq:model} are built from the same boson operators, and thus, they do not commute.  Consequently, spontaneous symmetry breaking in one species of the pseudo-spins can induce explicit symmetry breaking for the rest of the spin species, which results in the oscillatory behavior unless the explicitly broken symmetry is restored, see Appendix~\ref{sec:SSB}  for more details.

One can also restore the occupation of the four levels in the steady-state superradiant phase
\begin{equation}\label{eq:SRay_population}
    \begin{aligned}
& c_0^{\dagger} c_0=\frac{\mu\left(1-2 J^z\right)}{2} \\
& c_1^{\dagger} c_1=\frac{\mu\left(1+2 J^z\right)}{2} \\
& c_2^{\dagger} c_2=\frac{(1-\mu)\left(1-2 J^z\right)}{2} \\
& c_3^{\dagger} c_3=\frac{(1-\mu)\left(1+2 J_z\right)}{2},
\end{aligned}
\end{equation}
where $\mu$  is the fraction of atoms, initialized in state $\ket{0},$ $N_0= N c_0^{\dagger}c_0=\mu N.$

It is worth mapping the solution in terms of spins (or bosons) back to the  microscopic observables in the model~\eqref{eq:BEC}. In this way, one can unravel phase transition in the model~\eqref{eq:model} in the form of the self-organization transition(s) in terms of  atomic degrees of freedom, such as condensate density and magnetization. 

The density of the upper spinor component is $\rho_{\uparrow}=|\psi_{\uparrow}|^2=|\langle c_0\rangle\phi_0+\langle c_1\rangle \phi_1|^2=|\langle c_0\rangle|^2\phi_0^2+ (\langle c_0^\dagger \rangle\langle c_1\rangle+\langle c_1^{\dagger}\rangle \langle c_0\rangle)\phi_0\phi_1+|\langle c_1\rangle|^2\phi_1^2.$ The first term here is constant, while the second one takes the form $\cos kx\cos kz\propto \phi_1$ and describes the creation of the checkerboard lattice with the periodicity $2\pi/k$ when the system is in the SR phase, cf. Fig.~\ref{fig:level_schemes_Hs}(d,e). 
Similarly, the density of the lower spinor component reads $\rho_{\downarrow}=\left|\psi_{\downarrow}\right|^2=\left|\left\langle c_2\right\rangle \phi_0+\left\langle c_3\right\rangle \phi_1\right|^2.$
At the same time, one can calculate the spatial distribution of the spin through the lattice. The three components of this spin are given by  $\sigma^z=\left(|\psi_{\uparrow}|^2-|\psi_{\downarrow}|^2\right)/2,$
$\sigma_x=\Re\left(\psi_{\uparrow}^*\psi_{\downarrow}\right),$
$\sigma_y=\Im\left(\psi_{\uparrow}^*\psi_{\downarrow}\right).$   Here, we denote spin-1/2 operators with $\sigma$ to highlight the effective two internal spin levels nature of the effective model (operator $F$ above is defined on the spin-1 manifold). In the main text, we have plotted the values of these spins, calculated at the center of lattice cells using arrows on top of the distribution of the condensate density.

\subsection{$\eta_s=0$ case}

The model~\eqref{eq:hamiltonian_b} is a two-spin Dicke model with disorder in the level splittings. By deriving stationary solutions for this model, one can recover that
the critical coupling, at which transition to the SR phase occurs, depends on the initial state. The resulting expression is given by Eq.~\eqref{eq:critical_coupling}.
The $S^z$ spin components can be further found from
$\left[\left(\omega_{s}+\omega_{0}\right)+\frac{2\omega_{z}\eta ^{2}}{\omega_{z}^{2}+\kappa^{2}}S_{z}^{03}\right]S_{x}^{03}=0$
and
$\left[\left(\omega_{s}-\omega_{0}\right)+\frac{2\omega_{z}\eta ^{2}}{\omega_{z}^{2}+\kappa^{2}}S_{z}^{12}\right]S_{x}^{12}=0.$ Also, note that when one or both of the level splittings $\omega_0\pm \omega_s$ become negative, the initial state with particles prepared in the ground momentum state $\psi=\sqrt{\mu}\ket{0}+\sqrt{1-\mu}\ket{2}$ is effectively a population inverted state. Thus, the transition to the SR phase appears on longer timescales after the system relaxes to the ground state (corresponding to an excited momentum state). In this case, one can first observe decay with a `burst' of atoms to the excited momentum states $\ket{1} $ and $\ket{3},$ and then approach the correct SR state with a finite population of ground and excited momentum states. 

As it is illustrated in the main text, the SR transition in the spin $S,$ which is built simultaneously from different momentum and spin atomic states [cf. Eq.~\eqref{eq:levels}], can act as driving for both spin and density (momentum) of the BEC. Examples of such non-stationary behavior of internal/external degrees of freedom are shown in Fig.~\ref{fig:1}(d,e) and Fig.~\ref{fig:entanglement_lattice}(b-c).
This non-stationary behavior also can be seen by analyzing equations of motion for pseudo-spins $\mathcal T$ and $J:$
\begin{equation}
    \begin{aligned}
        \dfrac{dJ_{01}^{-}}{dt}	&=-2i\omega_{0}J_{01}^{-}+2i\eta \Re(a_{z})(\mathcal{T}_{13}^{+}-\mathcal{T}_{02}^{-})\\
\dfrac{dJ_{23}^{-}}{dt}	&=-2i\omega_{0}J_{23}^{-}+2i\eta \Re(a_{z})(\mathcal{T}_{13}^{-}-\mathcal{T}_{02}^{+})\\
\dfrac{d\mathcal{T}_{13}^{-}}{dt}	&=-2i\omega_{s}\mathcal{T}_{13}^{-}+2i\eta \Re(a_{z})(J_{23}^{-}-J_{01}^{+})
\\
\dfrac{d\mathcal{T}_{02}^{-}}{dt}	&=-2i\omega_{s}\mathcal{T}_{02}^{-}+2i\eta \Re(a_{z})(J_{23}^{+}-J_{01}^{-}).
\\
    \end{aligned}
\end{equation}
These equations result in the following precession frequencies
    
\begin{equation}
    \begin{aligned}
        \Omega&\to-\sqrt{2}\sqrt{-\frac{1}{4}\sqrt{\left(16\eta ^{2}\Re(a_{z})^{2}+4\omega_{0}^{2}+4\omega_{s}^{2}\right)^{2}-64\omega_{0}^{2}\omega_{s}^{2}}-4\eta ^{2}\Re(a_{z})^{2}-\omega_{0}^{2}-\omega_{s}^{2}}	\\
\Omega&\to-\sqrt{2}\sqrt{\frac{1}{4}\sqrt{\left(16\eta ^{2}\Re(a_{z})^{2}+4\omega_{0}^{2}+4\omega_{s}^{2}\right)^{2}-64\omega_{0}^{2}\omega_{s}^{2}}-4\eta ^{2}\Re(a_{z})^{2}-\omega_{0}^{2}-\omega_{s}^{2}}.
    \end{aligned}
\end{equation}

\section{\label{sec:Entanglement} Spin-momentum entanglement}
In this Appendix, we show how the von Neumann entropy witnesses correlations between spin and momentum.  Similar calculations for the negativity~\cite{negativity} and concurrence~\cite{wootters2001entanglement,Zou_2022} have also been performed. In our simulations, both quantities behaved similarly for all simulations, cf. Fig.~\ref{fig:quantifiers}.

\begin{figure} 
    \centering
    \includegraphics[width=0.75\linewidth]{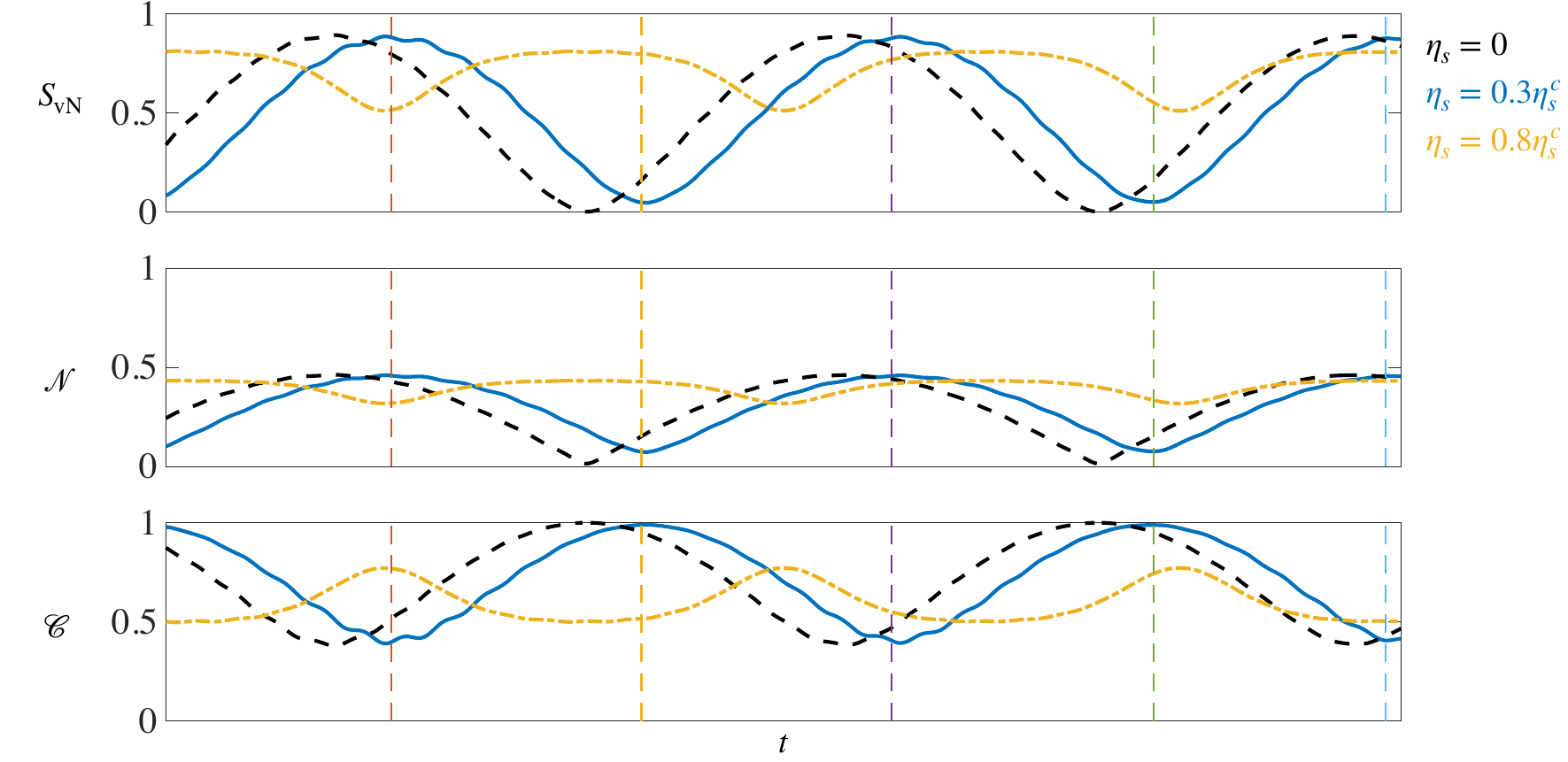}
    \caption{Dynamics of (a) entanglement entropy $S_{\operatorname{vN}}$, (b) negativity $\mathcal{N}$, and (c) concurrence $\mathscr{C}$ for parameters as in Fig.~\ref{fig:lattice_photon} in the main text. Different lines correspond to different values of coupling $\eta_s.$ \ox{The vertical dashed lines added to guide the eye and show the time moments when the entropy measures for $\eta_s=0.3\eta_s^c$ case take extremal values during dynamics.}}
    \label{fig:quantifiers}
\end{figure}

We can write down the atomic state as a superposition 
\begin{equation}    |\psi\rangle=\alpha|0\rangle+\beta|1\rangle+\gamma|2\rangle+\delta|3\rangle,
\end{equation}
where $\alpha=\langle c_0\rangle,$ $\beta=\langle c_1 \rangle,$ $\gamma=\langle c_2\rangle$ and $\delta=\langle c_3 \rangle, $ with 
 $|\alpha|^{2}+|\beta|^{2}+|\gamma|^{2}+|\delta|^{2}=1$ as the total number of atoms is conserved and normalized. Rewriting states $\ket{0},\ldots,\ket{3}$ in terms of spin and momentum states, see Eq.~\eqref{eq:levels}, the state of the system reads 
\begin{equation}
|\psi\rangle=\alpha|0\rangle_{m}\otimes|\downarrow\rangle_{s}+\beta|1\rangle_{m}\otimes|\downarrow\rangle_{s}+\gamma|0\rangle_{m}\otimes|\uparrow\rangle_{s}+\delta|1\rangle_{m}\otimes|\uparrow\rangle_{s}. 
\end{equation}
We now can construct a reduced density matrix by summing over the spin degree of freedom
\begin{equation}   \tilde{\rho}=\langle\mbox{spin}|\psi\rangle\langle\psi|\mbox{spin}\rangle
\end{equation}
or by summing over the momentum states 
\begin{equation}  \tilde{\rho}=\langle\mbox{momentum}|\psi\rangle\langle\psi|\mbox{momentum}\rangle.
\end{equation}
The reduced density matrix reads
\begin{equation}
\begin{aligned} \tilde{\rho}=\left[\begin{array}{ll}
|\alpha|^{2}+|\beta|^{2} & \alpha\gamma^{*}+\beta\delta^{*}\\
\alpha^{*}\gamma+\beta^{*}\delta & |\gamma|^{2}+|\delta|^{2}
\end{array}\right]\end{aligned}    
\end{equation}
and can be easily diagonalized.
The eigenvalues of this reduced density matrix are
$$\lambda=\dfrac{1}{2}\pm\dfrac{1}{2}\sqrt{1-4\left(|\alpha|^{2}|\delta|^{2}+|\beta|^{2}|\gamma|^{2}-\alpha\beta^{*}\gamma^{*}\delta-\alpha^{*}\beta\gamma\delta^{*}\right)}$$
or can alternatively be rewritten in terms of spins
$$\lambda=\dfrac{1}{2}\pm\dfrac{1}{2}\sqrt{1-4\left(S_{03}^{-}S_{03}^{+}+S_{12}^{-}S_{12}^{+}-J_{01}^{+}J_{23}^{-}-J_{01}^{-}J_{23}^{+}\right)}.$$
When $\eta =0,$ the system always remains in the pure state as the momentum state is separable in this case; in the superradiant phase, the fraction of atoms in the excited momentum state for $m_F=0$ is the same as the fraction of excited states for $m_F=1$.

On the other hand, when $\eta_{s}=0,$ the interaction $\eta $ induces entanglement between spin and momentum. The easiest way to see this is to consider the two-level case when all atoms are initially  prepared in state $\ket{0}.$ Then, by increasing the coupling $\eta $ above its critical value, the wave function of the state becomes 
$\psi=\alpha\ket{0}+\delta\ket{3}=\alpha |0\rangle_m \otimes|\downarrow\rangle_s+\delta |1\rangle_m \otimes|\uparrow\rangle_s,$
which is not separable in spin and momentum and, thus, is entangled. This state is maximally  entangled when $\alpha=\delta=1/\sqrt{2},$ and in terms of spin and momentum degrees of freedom, the state of the system is symmetric spin-momentum configuration.

\section{\label{sec:SSB} Spontaneous and explicit symmetry breaking}
In this Appendix, we gather arguments to elucidate the source of the non-equilibrium oscillatory phases that arise when one coupling surpasses the critical threshold while the other remains below it [cf. red and blue regions in the phase diagram in Fig.~\ref{fig:phase_diagram}(a)].

Let us consider the case when $\eta_{s}>\eta_{s}^{c},$ i.e., the cavity mode $a_y$ is the important mode. 
The SR transition in $H_s$ appears when the corresponding $\mathbb{Z}_{2}$ symmetry
of the model is broken. For the subsystem built on the momentum
states (via photon mode $a_y$, \cite{baumann2010dicke}), which is described by the Hamiltonian 
\begin{equation}
  H_{s}=\omega_{ {y}}a_{ {y}}^{\dagger}a_{ {y}}+\omega_{0}\left( {c}_{1}^{\dagger} {c}_{1}+ {c}_{3}^{\dagger} {c}_{3}\right)+\eta_{s}\left( {a}_{ {y}}+ {a}_{ {y}}^{\dagger}\right)\left( {c}_{0}^{\dagger} {c}_{1}+ {c}_{1}^{\dagger} {c}_{0}+ {c}_{2}^{\dagger} {c}_{3}+ {c}_{3}^{\dagger} {c}_{2}\right),  
\end{equation}
we have the condition
\begin{equation}
    \begin{aligned}
      a_{y}&\to-a_{y}\\
  \left( {c}_{0}^{\dagger} {c}_{1}+ {c}_{1}^{\dagger} {c}_{0}+ {c}_{2}^{\dagger} {c}_{3}+ {c}_{3}^{\dagger} {c}_{2}\right)&\to-\left( {c}_{0}^{\dagger} {c}_{1}+ {c}_{1}^{\dagger} {c}_{0}+ {c}_{2}^{\dagger} {c}_{3}+ {c}_{3}^{\dagger} {c}_{2}\right).  
    \end{aligned}
\end{equation}
The second line above corresponds to $J^{x}\to-J^{x}$ for the Dicke model. Under the corresponding transformation in Eq.~\eqref{eq:transformation_main} 
the bosonic part of the Hamiltonian transforms as 
\begin{equation}
    \begin{aligned}
        & \left( {c}_{0}^{\dagger} {c}_{1}+ {c}_{1}^{\dagger} {c}_{0}+ {c}_{2}^{\dagger} {c}_{3}+ {c}_{3}^{\dagger} {c}_{2}\right)\to\left( {c}_{0}^{\dagger} {c}_{1}e^{i\left(\phi_{0}-\phi_{1}\right)}+ {c}_{1}^{\dagger} {c}_{0}e^{i\left(\phi_{1}-\phi_{0}\right)}+ {c}_{2}^{\dagger} {c}_{3}e^{i\left(\phi_{2}-\phi_{3}\right)}+ {c}_{3}^{\dagger} {c}_{2}e^{i\left(\phi_{3}-\phi_{2}\right)}\right)\\
 & =-\left( {c}_{0}^{\dagger} {c}_{1}+ {c}_{1}^{\dagger} {c}_{0}+ {c}_{2}^{\dagger} {c}_{3}+ {c}_{3}^{\dagger} {c}_{2}\right),
    \end{aligned}
\end{equation}
thus seting the constrains  $e^{\pm i\left(\phi_{1}-\phi_{0}\right)}=-1$ and $e^{\pm i\left(\phi_{2}-\phi_{3}\right)}=-1$.
These constraints bring us to the following condition on the relative phases:
\begin{equation}
    \begin{aligned}
        \phi_{0}-\phi_{1} & =\pi\pm2\pi n\\
\phi_{2}-\phi_{3} & =\pi\pm2\pi m.
    \end{aligned}
\end{equation}
As only the relative phase between two bosonic fields enters the
Hamiltonian, we get two conditions for four phases.
If we perform the same transformation on the second interacting term in the {total} Hamiltonian in Eq.~\eqref{eq:model}, which reads
\begin{equation}    \eta\left( {a}_z+ {a}_z^{\dagger}\right)\left( {c}_1^{\dagger}  {c}_2+ {c}_0^{\dagger}  {c}_3+ {c}_2^{\dagger}  {c}_1+ {c}_3^{\dagger}  {c}_0\right),
\end{equation}
we find that it induces an additional phase for photon
field $a_{z},$ which provokes an explicit symmetry breaking
\begin{equation}
    \begin{aligned}
        & \eta \left( {a}_{ {z}}e^{-i\phi_{z}}+ {a}_{ {z}}^{\dagger}e^{i\phi_{z}}\right)\left( {c}_{1}^{\dagger} {c}_{2}e^{i\left(\phi_{1}-\phi_{2}\right)}+ {c}_{0}^{\dagger} {c}_{3}e^{i\left(\phi_{0}-\phi_{3}\right)}+ {c}_{2}^{\dagger} {c}_{1}e^{-i\left(\phi_{1}-\phi_{2}\right)}+ {c}_{3}^{\dagger} {c}_{0}e^{-i\left(\phi_{0}-\phi_{3}\right)}\right)=\\
 & =\eta \left( {a}_{ {z}}e^{-i\phi_{z}}+ {a}_{ {z}}^{\dagger}e^{i\phi_{z}}\right)\left( {c}_{1}^{\dagger} {c}_{2}e^{i\left(\phi_{1}-\phi_{3}-\pi\right)}+ {c}_{0}^{\dagger} {c}_{3}e^{i\left(\phi_{1}+\pi-\phi_{3}\right)}+ {c}_{2}^{\dagger} {c}_{1}e^{-i\left(\phi_{1}-\phi_{3}-\pi\right)}+ {c}_{3}^{\dagger} {c}_{0}e^{-i\left(\phi_{1}+\pi-\phi_{3}\right)}\right)=\\
 & =-\eta \left( {a}_{ {z}}e^{-i\phi_{z}}+ {a}_{ {z}}^{\dagger}e^{i\phi_{z}}\right)\left( {c}_{1}^{\dagger} {c}_{2}e^{i\left(\phi_{1}-\phi_{3}\right)}+ {c}_{0}^{\dagger} {c}_{3}e^{i\left(\phi_{1}-\phi_{3}\right)}+ {c}_{2}^{\dagger} {c}_{1}e^{-i\left(\phi_{1}-\phi_{3}\right)}+ {c}_{3}^{\dagger} {c}_{0}e^{-i\left(\phi_{1}-\phi_{3}\right)}\right).
    \end{aligned}
\end{equation}
One can recognize that the bosonic part gains a phase $\pm\left(\phi_{1}-\phi_{3}\right)$
which,  generally, can take an arbitrary value \ox{and cannot be immediately compensated by the phase $\phi_z$}. So, as soon as we turn
on coupling $\eta $, we break the symmetry of the Hamiltonian.
In general, to restore the symmetry at finite time $T$, the population of \ox{the two levels (one excited and another ground momentum states) must reach zero value. In practice, when $\omega_s>0$, particles from levels $\ket{2}$ and $\ket{3}$ drift to states $\ket{0}$ and $\ket{1}$ (or in the opposite direction when $\omega_s<0$). }

The breaking of the symmetry can be seen also in the following calculation. Given the steady state for
 $\eta_{s}>\eta_{s}^{c}$ and $\eta =0$ [cf. Eq.~\eqref{eq:SRay_population}], 
after quenching $\eta $, 
the photon mode $a_{z}$ becomes
\begin{equation}
   a_{z}=-2\dfrac{\eta \left(\omega_{z}-i\kappa\right)}{\left(\omega_{z}^{2}+\kappa^{2}\right)}\left(\sqrt{(1-\mu)\mu(1+2J^{z})\left(1-2J^{z}\right)}\right) 
\end{equation}
which is non-zero for  $\mu\ne0,1$. Here, the typical time at which the photon approaches the value above is $\approx \text{1/\ensuremath{\kappa}}$ negligibly small.
The explicit breaking of the symmetry in the term 
\begin{equation}\label{eq:explsym}    
\begin{aligned}
&\eta\left( {a}_z+ {a}_z^{\dagger}\right)\left( {c}_1^{\dagger}  {c}_2+ {c}_0^{\dagger}  {c}_3+ {c}_2^{\dagger}  {c}_1+ {c}_3^{\dagger}  {c}_0\right)\to \\
&-\eta \left( {a}_{ {z}}e^{-i\phi_{z}}+ {a}_{ {z}}^{\dagger}e^{i\phi_{z}}\right)\left( {c}_{1}^{\dagger} {c}_{2}e^{i\left(\phi_{1}-\phi_{3}\right)}+ {c}_{0}^{\dagger} {c}_{3}e^{i\left(\phi_{1}-\phi_{3}\right)}+ {c}_{2}^{\dagger} {c}_{1}e^{-i\left(\phi_{1}-\phi_{3}\right)}+ {c}_{3}^{\dagger} {c}_{0}e^{-i\left(\phi_{1}-\phi_{3}\right)}\right)
\end{aligned}
\end{equation}
takes the system out of equilibrium. To reach  steady state, the Hamiltonian must regain the symmetry via one of two mechanisms:
\begin{enumerate}
    \item  The decay of the photon number of the auxiliary mode to zero, $n_z=0$. 
    \item The phases of the bosons adjusting to satisfy the condition $\phi_1-\phi_3\equiv 0.$
\end{enumerate}
The oscillatory phase accompanies the system's dynamics until one of these two possibilities is realized. 
Below, we discuss both scenarios in more detail.


\section{\label{sec:atom_only} Relaxation dynamics at late times}
In this Appendix, we study the properties of the system at late times and evaluate the time $T$ at which the system restores its symmetry.

 To restore the explicitly broken symmetry
 one needs to 
 tune  the emergent phases in front of the auxiliary photons, $\phi_c$, and  pseudo-spins, $\phi_i-\phi_j$, to zero. 
 This procedure  can be trivially done  when the corresponding occupation of the photon or atomic levels equals zero 
 due to the ambiguity of the phase of these complex numbers when their magnitude is zero. 
 In the former case, the symmetry restoration occurs asymptotically at time $T\to\infty$ through the slow decay of the cavity photon magnitude induced by the dissipation $\kappa$. In line with this scenario, the dynamics, when expressed in terms of the two cavity fields, resemble the behavior depicted in the middle panels of Fig.~\ref{fig:phase_diagram}(c-d) during the course of the experiment.
 
  According to the second mechanism, which takes a finite time $T$, during dynamics all particles slowly transfer to the lowest energy pair of ground and excited momentum states. For instance,  if $\omega_s>0,$ all particles will redistribute to levels $\ket{0} $ and $\ket{1}$ resulting into $\left\langle c_2(T)\right\rangle=0$ and $\left\langle c_3(T)\right\rangle=0$. Then, when states with higher energy become empty, the explicitly broken symmetry of the system can be restored. Eventually, the phases of both cavity fields remain constant, and both fields start approaching true superradiant states with $n_y\ne 0$ and $n_z\ne 0$, cf. the right panels in Fig.~\ref{fig:phase_diagram}(c-d).

\begin{figure}
    \centering
    \includegraphics[width=0.7\linewidth]{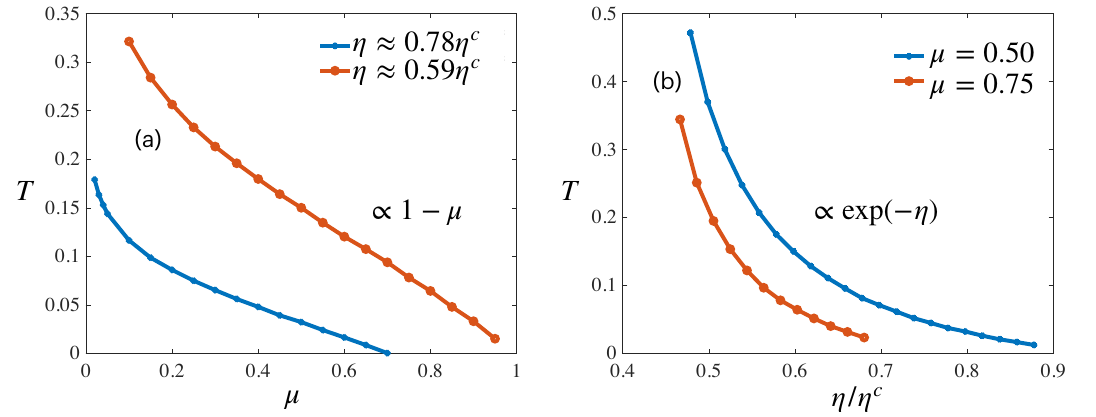}
    \caption{{Symmetry restoration time} $T$ (in seconds) as a function of (a) the initial population of the level $\ket{0}$, i.e., $N_0=\mu N$ and (b) of the photon-matter coupling to the auxiliary cavity field. Here we consider mode $a_y$ as the main and mode $a_z $ as the auxiliary. Here, for simplicity, we set $\omega_y=\omega_z\approx\kappa/3,$ $\omega_s=0.2.$ For these two simulations, we prepare the system in the superradiant phase for $\eta_s=1.2\eta_s^c,$ and then perform a fast ramp of $\eta<\eta^c.$}
    \label{fig:T}
\end{figure}

 Fig.~\ref{fig:T} shows the scaling of the relaxation time  with (a) the fraction of particles at $t=0$ in state $\ket{0},$ and (b) with the strength of the smaller of the two couplings, i.e, auxiliary one. \ox{Here, we prepare the system in the superradiant steady state with $\eta_s>\eta_s^c$, and then rapidly ramp the second coupling $\eta$. In these simulations we consider the mode $a_z$ as the auxiliary one for simplicity  because its critical coupling depends on $\mu$.} 
 The finite time $T$ corresponds to the scenario when symmetry restoration takes place through the transferring of all particles to the lower energy pair of ground and excited momentum states.
 For instance, for $\omega_s>0$, the more particles are initially prepared in state $\ket{2},$ the longer is $T$ [cf. Fig.~\ref{fig:T}(a)]. Assuming that the speed of particles transferring from levels $\ket{2}$ and $\ket{3}$ remains constant and depends solely on the couplings and detunings, the relaxation time will be proportional to the population of the zero momentum state with the higher energy. One would estimate in this case 
 $ T \propto (1-\mu) $ as we confirm numerically in panel (a). 
 At the same time, when $\omega_s<0,$ the pair of states that have smaller energy are $\ket{2}$ and $\ket{1}, $ and the relaxation time scales like $T\propto\mu.$ This can also be seen from the Hamiltonian $H_{\operatorname{tot}}$ which in the normal state is invariant under the transformation $\mu\leftrightarrow1-\mu$ when $\omega_s\to-\omega_s.$
 
 On the other hand, the symmetry restoration time scales with the coupling to the auxiliary mode [$\eta $ in Fig.~\ref{fig:T}(b)] like $T\propto \exp (-\eta ).$  It means that the closer this auxiliary coupling is to its critical value, the faster the symmetry is restored. Consequently, operating within a parameter regime where the coupling to the auxiliary mode remains significantly below the critical threshold, yet remains finite, gives us the opportunity to probe the atomic dynamics within the system with minimal disruption.

Below we consider a few fine-tuned limits in which a stationary state can not be reached within the operational timescales of the experiment.

\subsection{$\kappa/\omega_c\to \infty$ limit}

The relaxation time $T$ depends on the strength of the photon-matter coupling with the auxiliary mode; namely, the stronger the interaction, the faster the relaxation [see Fig.~\ref{fig:T}(b)]. On the other hand, the strength of this coupling controls the population of the auxiliary cavity mode,   $n\propto \eta^2/(\omega_c^2+\kappa^2).$ As such, it is instructive to explore  the regime where $\eta/\sqrt{\omega_c^2+\kappa^2}$ is large enough to produce sufficiently large cavity field amplitude to enable detections of oscillations in the auxiliary cavity field. We would also like to simultaneously satisfy the condition $\eta\ll\eta_c$ so that the system's restoration of its symmetry becomes protracted to exponentially late times. 
These conditions can be satisfied simultaneously when $\omega_c\ll \kappa$, a regime in which the critical coupling diverges like $\sqrt{\omega_0\kappa^2/\omega_c}$. The extreme case of $\omega_c=0$ corresponds to the critical coupling tending to infinity. 
Counter-intuitively, in this case, one enters  a strong dissipation regime which  prevents the relaxation.

The main idea of why the strong dissipation regime prevents relaxation can be explained as follows: according to the Hamiltonian $H_{\operatorname{tot}}$ [Eq.~\eqref{eq:model}], pseudo-spins are coupled to the real quadrature of the cavity fields, $a+a^{\dagger}$. If  the cavity fields are coupled to the $x-$components of pseudo-spins, then the cavity field is imaginary, $a_y\propto 2  J^x/(-i\kappa)$ (or $a_z\propto 2 S^x/(-i\kappa)$).  Here  the cavity decay rate causes a phase shift  $\phi_{\kappa}^{y,z}=\tan ^{-1}\left(-\kappa / \omega_{y,z}\right)$ of the field scattered into the cavity by the atomic system~\cite{dogra2019dissipation}. Thus, by setting the cavity detuning $\omega_y$ or $\omega_z$ to be much smaller than $\kappa$, one can make the corresponding cavity field imaginary and eliminate its feedback on the spin dynamics. For instance, when  $\eta_s>\eta_s^c,$ $\eta <\eta ^c$ and $\omega_y\ne 0,$ $\omega_z\to 0$ [blue region in Fig.~\ref{fig:phase_diagram}(a)], $a_y$ 
approaches its steady state value, while 
$a_z\approx -2 \eta \left(S_{03}^x+S_{12}^x\right) /\left(-i \kappa\right) $ oscillates together with the 
precession of the pseudo-spin $S^x.$ However, as $(a_z+a_z^{\dagger})\to 0,$ the non-stationary behavior of the cavity field does not impact the dynamics of pseudo-spin $S^x$.
In other words, subsystem $H_s$ constantly induces precession for pseudo-spin $S^x$ (and thus oscillations to the cavity mode $a_z$) due to the explicit symmetry breaking; it does not, however, experience feedback from subsystem $H.$  As a result, the lack of  reciprocal interaction between the two subsystems prevents the system from reaching a steady state, and oscillations in the auxiliary cavity mode $a_z$ survive for an arbitrarily long time. 

When $\eta_{s}<\eta_s^c$ and $\eta >\eta ^c$ [red region in the phase diagram in Fig.~\ref{fig:phase_diagram}(a)], one should set $\omega_y\ll\kappa$ to prevent the system from reaching its steady state. In this case, the cavity mode $a_y$ will exhibit oscillations around zero for an arbitrarily long time, reflecting the precession of pseudo-spin $J^x$. The cavity mode $a_z$, in turn, will approach its steady state value, determined solely by the Hamiltonian $H$ [Eq.~\eqref{eq:hamiltonian_b}] and initial conditions.

Interestingly, in the opposite limit where $\omega_c\gg \kappa$, the system can also experience slow relaxation. The critical coupling scales like  $\eta_c\propto \sqrt{\omega_0\omega_c}\to \infty$; it is easier to keep the coupling strongly subcritical while still large enough to enable read-out.  However, as  the cavity occupation scales as $n\propto 1/\omega_c^2$, in order to keep $n$ non-zero, it is essential to keep the cavity detuning finite.

\subsection{$\omega_s\to 0$ limit} 

 One can recognize from the level scheme in Fig.~\ref{fig:1}(b) that when $\omega_s=0,$ the two ground states in the momentum variables, $\ket{0},\ket{2}$, and the two excited ones $\ket{1},\ket{3}$ are degenerate. In this case, the Hamiltonian acquires an additional symmetry under exchange between ground or excited states, namely $c_1\leftrightarrow c_3,$ $c_0\leftrightarrow c_2$ (see Hamiltonian in the bosonic representation in Appendix~\ref{sec:derivation}). As a consequence, the dynamics for the pair of fields $c_{1}, c_{3}$ (and likewise for $c_{0},c_{2}$) occur at the same frequencies $c_{1},c_{3}\propto\exp(i\Omega t).$ Accordingly, the phase that explicitly breaks the symmetry of the Hamiltonian, $\pm(\phi_{1}-\phi_{3}),$  remains constant over time,  determined solely by the initial conditions (which can be arbitrary and are not restricted in general).  Thus, after the quench, both spins $J^x$ and $S^x$ gain fixed time values, and one can observe superradiance in both cavity modes simultaneously. Interestingly,  in this case, $n_y$ and $n_z$ do not oscillate over time, and non-stationary behavior can only be observed  at the level of the  atomic observables. In particular, particles redistribute between different excited or ground momentum states so that the overall number of particles in the upper and lower spinor components oscillates over time around a common time-averaged value. 

 \section{Three-level model\label{sec:three_levels}}

We now revisit the possibility of probing the system's dynamics with the auxiliary  cavity field in a three-level system. As mentioned in the main text, such a model includes single ground momentum state $\ket{2}$, and two excited momentum states $\ket{1},$ $\ket{3},$ where we keep notation as in the Eq.~\eqref{eq:levels}. Such Hamiltonian can be implemented when fixing $\mathbf{e}_p=\mathbf{e}_x$ in Eq.~\eqref{eq:classical_drive} and considering resonant spin changing and spin-dependent processes, cf. implementations in Refs.~\cite{dogra2019dissipation,Ferri}. Here we omit the implementation of the three-level model, concentrating mostly on the physical phenomena, compared to the four-level model in Eq.~\eqref{eq:model}. 
To do so, we study the effect of spontaneous symmetry breaking in one sector of the Hamiltonian on the dynamical properties in another sector. As we demonstrate below, explicit symmetry breaking in the self-ordered phase(s) is not pronounced in the three-level case, thereby resulting in trivial system dynamics. 

The Hamiltonian of our three-level model reads
    \begin{equation}
\begin{aligned}
	H_3	&=\omega_{ {y}}a_{ {y}}^{\dagger}a_{ {y}}+\omega_{ {z}}a_{ {z}}^{\dagger}a_{ {z}}+\omega_0\left( {c}_{1}^{\dagger} {c}_{1}+ {c}_{3}^{\dagger} {c}_{3}\right)+\omega_{s}( {c}_{2}^{\dagger} {c}_{2}+ {c}_{3}^{\dagger} {c}_{3})+\\
	&+\eta \left( {a}_{ {z}}+ {a}_{ {z}}^{\dagger}\right)\left( {c}_{1}^{\dagger} {c}_{2}+ {c}_{2}^{\dagger} {c}_{1}\right)
 +\eta_{s}\left( {a}_{ {y}}+ {a}_{ {y}}^{\dagger}\right)\left( {c}_{2}^{\dagger} {c}_{3}+ {c}_{3}^{\dagger} {c}_{2}\right).
\end{aligned}
\end{equation}
Let us consider the effect of the spontaneous symmetry breaking in one subsystem on the dynamics of another, similarly as it is done in Appendix~\ref{sec:SSB}.
Firstly, we consider a steady  state when $\eta_s>\eta_s^c$ and $\eta =0$. In this case, the symmetry of a subsystem involving photon mode $a_y$ is broken, meaning there exist two solutions, satisfying 
\begin{equation}
\begin{aligned}
a_y & \rightarrow-a_y \\
\left( {c}_2^{\dagger}  {c}_3+ {c}_3^{\dagger}  {c}_2\right) & \rightarrow-\left( {c}_2^{\dagger}  {c}_3+ {c}_3^{\dagger}  {c}_2\right).
\end{aligned}
\end{equation}
Applying transformation~\eqref{eq:transformation_main}, we can see that such  transformation 
\begin{equation}
    \begin{aligned}         \left( {c}_{2}^{\dagger} {c}_{3}+ {c}_{3}^{\dagger} {c}_{2}\right)&\to\left( {c}_{2}^{\dagger} {c}_{3}e^{i\left(\phi_{2}-\phi_{3}\right)}+ {c}_{3}^{\dagger} {c}_{2}e^{i\left(\phi_{3}-\phi_{2}\right)}\right)\\
 & =-\left( {c}_{2}^{\dagger} {c}_{3}+ {c}_{3}^{\dagger} {c}_{2}\right)
    \end{aligned}
\end{equation}
 sets the following constraints on relative phases
\begin{equation}
 \phi_{2}-\phi_{3}=\pi\pm 2\pi m.   
\end{equation}
Note that there are no restrictions on the phase $\phi_{1}$ because photon mode $a_y$ is not coupled to the level $\ket{1}$. As such, making the second coupling, $\eta$, non-zero does not induce explicit symmetry breaking.  
This fact can be seen by  applying the transformation~\eqref{eq:transformation_main} to this term: 
\begin{equation}
   \left( {a}_{ {z}}e^{-i\phi_{z}}+ {a}_{ {z}}^{\dagger}e^{i\phi_{z}}\right)\left( {c}_{1}^{\dagger} {c}_{2}e^{i\left(\phi_{1}-\phi_{2}\right)}+ {c}_{2}^{\dagger} {c}_{1}e^{-i\left(\phi_{1}-\phi_{2}\right)}\right) \propto \langle c_1\rangle  \equiv 0.
\end{equation}
The phase of the boson field $c_1$ can always compensate for the restricted phase of $c_2$. Note that we assume that level $\ket{1}$ is unoccupied before increasing $\eta$, and thus the phase of $c_1$ can be changed arbitrarily. 
The system is therefore stable against quenches $\eta <\eta ^c$. This behavior arises from the fact that each photon mode is  not coupled to all atomic levels; symmetry breaking in one interaction term does not imply explicit symmetry breaking in another.
More precisely, when $\eta_{s}>\eta_{s}^{c}$
and $\eta =0$, one finds that
\begin{equation}
    \begin{aligned}
     c_{1}^{\dagger}c_{1} & =0\\
c_{2}^{\dagger}c_{2} & =\frac{\left(1-2J^{z}\right)}{2}\\
c_{3}^{\dagger}c_{3} & =\frac{\left(1+2J_{z}\right)}{2}   
    \end{aligned}
\end{equation}
and after the quench of $\eta $, we get $a_{z}\propto\left(c_{1}^{\dagger}c_{2}+c_{2}^{\dagger}c_{1}\right)=0$ and the subsystem remains in steady state as long as $\eta /\eta ^c<\eta_s/\eta_s^c$.
The absence of restrictions on phase $\phi_{1}$ makes it impossible to induce competing conditions and push the subsystem out of equilibrium. Indeed, in the case of four levels, one cannot  manipulate the phase of a single boson separately, thereby resulting in the existence of long-lived oscillations of the auxiliary cavity field and the precession of corresponding pseudo-spins.

 \end{widetext}
\bibliography{ETH}

\end{document}